\documentclass[aps,prd,superscriptaddress,showpacs,twocolumn,
nofootinbib,eqsecnum]{revtex4}

\usepackage{amsmath,amsfonts}

\newcommand{\rhoc}{{\rho_*}}
\newcommand{\wc}{{w}}
\newcommand{\pc}{{p_*}}
\newcommand{\hc}{{h_*}}
\newcommand{\gammac}{{\gamma_*}}
\newcommand{\Uc}{{U_*}}
\newcommand{\dx}{{d^3\mathbf{x}}}
\newcommand{\cO}{{\cal O}}

\newcommand{\epsx}{\varepsilon_{x'}}
\newcommand{\Epsilon}{{\epsilon}}
\newcommand{\Uic}{{U_*}}
\newcommand{\Di}{{D}}

\newcommand{\Apot}{{A_*}}

\newcommand{\integ}{{\int \!}}

\allowdisplaybreaks

\begin{document}

\author{Guillaume Faye}
\email{G.Faye@tpi.uni-jena.de}
\affiliation{Theoretisch-Physikalisches Institut,
Friedrich-Schiller-Universit\"at, Max-Wien-Platz 1, 07743 Jena, Germany}

\author{Gerhard Sch\"afer}
\email{G.Schaefer@tpi.uni-jena.de}
\affiliation{Theoretisch-Physikalisches Institut,
Friedrich-Schiller-Universit\"at, Max-Wien-Platz 1, 07743 Jena, Germany}

\title{Optimizing the third-and-a-half post-Newtonian gravitational 
radiation-reaction force for numerical simulations}

\date{\today}

\pacs{04.25.-g}

\begin{abstract}
The gravitational radiation-reaction force acting on perfect fluids at $3.5$
post-Newtonian order is cast into a form which is directly applicable to
numerical simulations. Extensive use is made of metric-coefficient changes
induced by functional coordinate transformations, of the continuity
equation, as well as of the equations of motion. We also present an expression
appropriate for numerical simulations of the radiation field causing
the worked out reaction force.
\end{abstract}

\maketitle

\section{Introduction}
The reaction force acting on isolated perfect fluids due to gravitational
radiation emission has been expressed by means of Burke-Thorne-like potentials
up to the $7/2$ post-Newtonian ($3.5$ PN) order, which corresponds to the
seventh order in power of the inverse of the speed of light $c$ \cite{B97}.
However, this particular form is apparently not convenient for computational
purposes as it involves up to the seventh time derivative of multipole
moments. A similar problem happens with the 1 PN radiation field which causes
the $3.5$ PN reaction force, since up to four time derivatives do apply
therein.

It is well known, e.g. see \cite{S83}, that the order of time derivatives in
the $2.5$ PN reaction force crucially depends on the chosen coordinate system.
The influence of general changes of coordinates onto the metric coefficients
of a many particle system has been worked out in e.g., \cite{DS85}, to the 2
PN order.  In generalizing this result to perfect fluids, combining it with
the choice of a suitable set of variables, we shall be able to reduce the
$3.5$ PN reaction force to a fifth time derivative object. Furthermore,
applying the continuity equation and the equations of motion will result in
expressions where only one time derivative remains left, which poses no
problems for a numerical implementation. In this way we generalize a previous
work by Rezzolla \emph{et al.} \cite{RSABS99} to the case of nonzero
mass-multipole moments. We are then able to give the complete set of
equations describing the fluid evolution up to $1$ PN $+~3.5$ PN order in a
similar form as in a former paper by Blanchet, Damour and Sch\"afer
\cite{BDS90}. In addition, we present the full explicit expression for the 1
PN radiation field, adapted to numerical simulations.

Though the formalism we propose limits to the case of adiabatic fluids, it may
still be used in a large range of astrophysical applications. Particularly it
provides a natural way to generalize the simulations of Oohara and Nakamura on
the coalescence of binary neutron stars \cite{ON97} achieved at the $1$ PN
conservative and $2.5$ PN dissipative levels, by adding the $3.5$ PN
contributions to the gravitational reaction force. Our equations are also
appropriate to study the effect of the gravitational damping, including the
mass quadrupole \cite{JAG03}, the mass octupole, as well as the current
quadrupole (treated separately in paper \cite{RSABS99}), on the evolution of
fluid modes in rotating compact stars \cite{S98} in the case where the fluid
viscosity is neglected. The role played by the bulk viscosity in reducing the
Chandrasekhar-Friedman-Schutz instability \cite{C70,FS78} may actually be
significant, but it can be temporarily ignored, regarding the complexity of
the problem. Our model of matter will eventually be improved in future works.
Let us point out finally that the adiabaticity condition only means the
entropy conservation of fluid particles along their trajectories, and does not
impose any specific relation between energy, entropy and pressure. The
equation of state linking these three variables
can thus be freely chosen. Suitable choices \cite{DFM02a,S02} permit to model
for instance
rather realistic stellar collapse processes within the present formalism.

\section{Effect of a gauge transformation on a metric} 
\label{sec:transformation}

In order to determine the coordinate system that minimizes the order of time
derivatives appearing in the reaction force, we need to know the specific
effect of a general change of coordinates on the functional form of the
metric. More precisely, we must extend the ``contact transformation''
investigated in paper \cite{DS85} for systems of point-like particles to the
case where the gravitational field is generated by a continuous distribution
of matter. To determine the order at which we need to operate, we notice that
the parts of the metric we are interested in contribute to the $1$ PN
acceleration, as well as to the $2.5$ and $3.5$ PN dissipative dynamics.
Products of terms corresponding to the $1$ PN and $2.5$ PN level in the
equations of motion cannot be neglected since they affect the $3.5$ PN
evolution. Therefore, the gauge transformation must be \emph{a priori}
investigated up to the quadratic order. However, we shall see that the linear
approximation is sufficient provided the difference between the old and new
coordinates is given in terms of the new variables. Still, some formulas, like
formula \eqref{eq:epsilon_square} which shows how two successive gauge
transformations of order $1$ PN and $2.5$ $+~3.5$ differs from the
transformation induced by the sum of the associated change of coordinates, can
only be obtained within the quadratic approximation. Now, rather than limiting
ourselves to the quadratic level, we shall give for completeness most of the
relations presented in this section to arbitrary high orders. The resulting
expressions will be longer, but more general and not fundamentally more
complicated. They may be useful for future works, although not strictly
required in the present paper. The reader solely interested in the application
to the actual problem of eliminating time derivatives from the $3.5$ PN
reaction force may skip this section. The reader exclusively interested in the
final result may go directly to the end of section \ref{sec:reaction}
(equations \eqref{eq:FS_equations}).

In a given coordinate grid, the metric $g_{\mu\nu}$ (Greek indices run from 0
to 3, and Latin indices from 1 to 3) is a function of the coordinate $x^\alpha
= (c t =x^0 , \mathbf{x} = x^i)$ and depends functionally on certain matter
fields $X_1, X_2, \ldots, X_k, \ldots$ In the case of barotropic fluids, for
instance, they can be the mass density $\rho$ and the 4-velocity field $u^\mu$
normalized to unity ($u^\mu u_\mu = -1$), or the baryonic mass density
$\rho_*$ and the linear 3-momentum density $M_i$, or other sets of relevant
variables. We have:
\[g_{\mu\nu} = g_{\mu\nu}(x^\alpha,X_1(y^\alpha), X_2(y^\alpha), \ldots,
X_k(y^\alpha)) \, ;\]
in short form, $g_{\mu\nu} = g_{\mu\nu}(x^\alpha, X_A(y^\alpha))$.
In our notation, the variables $x^\alpha$ contain the entire dependence on the
coordinates, whereas $y^\alpha$ are seen as mere ``dummy'' quantities.  
The fields $X_A(y^\alpha)$ are in fact themselves functionals of the
components of the stress energy-tensor and of the metric,
$X_A(y^\alpha) = F_A (y^\alpha, T^{\alpha\beta}, g_{\alpha\beta})$,
even if it is not indicated explicitly for simplicity. The
important point is that a coordinate transformation, which is passive
by essence, affects the function $X_A=X_A(y^\alpha)$.

Let us consider now the (exact) change of coordinates
\[ x^\alpha = x'^\alpha + \epsx^\alpha 
(x'^\beta,X'_A(y^\beta)) \, ,\] 
where $\epsx^\alpha$ is a function of $x'^\beta$ and a functional of
the fields $X'_A(y^\beta)$.
Such a transformation acts simultaneously on the coordinates, on
the metric and on the fields $X_A(y^\alpha)$. As a consequence, the new
components of $g'_{\mu\nu}$ are related to the old ones by
\begin{align} \label{eq:transformation_g}
g'_{\mu\nu}(x'^\alpha, X'_A (y^\alpha &)) = g_{\mu\nu}(x^\alpha, X_A(y^\alpha))
\nonumber \\ 
& +2 g_{\lambda(\mu} (x^{\underline{\alpha}},
X_A(y^{\underline{\alpha}}))\partial'_{\nu)} 
\epsx^\lambda(x'^\beta, X'_A(y^\beta)) \nonumber \\ +&~
g_{\lambda\rho}(x^\alpha, X_A(y^\alpha)) 
\partial'_\mu \epsx^\lambda (x'^\beta,
X'_A(y^\beta))  \nonumber \\ & \times \partial'_\nu \epsx^\rho (x'^\gamma,
X'_A(y^\gamma)) \, ,
\end{align}
where $\partial'_\lambda$ represents the
partial derivative with respect to $x'^\lambda$, while the brackets around
non-underlined indices mean their symmetrization. 
The fields $X'_A(y^\alpha)$ entering the argument of $g'_{\mu\nu}$ read
$X'_A(y^\alpha) = F_A(y^\alpha,T'^{\alpha\beta},g'_{\alpha\beta})$. Since the
dependence on $T'^{\alpha\beta}(z^\gamma)$ and $g'_{\alpha\beta}(z^\gamma)$ is
purely functional, 
the position variables $z^\gamma$ appearing there must be regarded as dummy,
but their other arguments are combinations of tensor components referring to
the current coordinate system:
$T'^{\alpha\beta} = T'^{\alpha\beta}(z^\gamma,g'_{\gamma\delta},\ldots)$ and 
$g'_{\alpha\beta} = g'_{\alpha\beta}(z^\gamma,X'_B)$. In
the post-Newtonian framework, the computation of $g'_{\mu\nu}$ is actually
restricted to a finite order in powers of $\epsx^\alpha$ if we assume
that the latter quantity 
is of order of the inverse of the speed of light $c$ or higher, namely 
$\epsx^\alpha =\cO(1/c)$. Therefore, when the function
$g_{\mu\nu}(x^\alpha, X_A(y^\alpha))$ is known, the
dependence of $g'_{\mu\nu}$ on $x'^\alpha$ can be obtained explicitly by
expanding the right-hand side of equation \eqref{eq:transformation_g} in powers
of $\epsx^\alpha = x^\alpha - x'^\alpha$ within the required
precision. Since the order of expansion can be arbitrarily high, we may
symbolically write
\begin{widetext}
\begin{align} \label{eq:dvpt_epsilon}
g'_{\mu\nu}(x'^\alpha, X'_A(y^\alpha)) =&~
g_{\mu\nu}(x'^\alpha, X_A(y^\alpha)) + \epsx^\lambda(x'^\alpha,
X'_A(y^\alpha)) \partial'_\lambda
g_{\mu\nu}(x'^\beta, X_A(y^\beta)) + 2 g_{\lambda (\mu}
(x'^{\underline{\alpha}}, X_A(y^{\underline{\alpha}})) \partial'_{\nu)} 
\epsx^\lambda(x'^\beta, X'_A(y^\beta)) \nonumber \\ &+
\sum_{k=0}^{+\infty} \frac{1}{k!} \bigg(
\frac{\epsx^{\lambda_1 \ldots \lambda_k \lambda \rho}}{(k+1)(k+2)} (x'^\alpha,
X'_A(y^\alpha)) \partial'_{\lambda_1 \ldots \lambda_k \lambda \rho}
g_{\mu\nu}(x'^\beta, X_A(y^\beta)) \nonumber \\& + \frac{2}{k+1} 
\epsx^{\lambda_1 \ldots \lambda_k \rho}(x'^\alpha, X'_A(y^\alpha))
\partial'_{\lambda_1 \ldots 
\lambda_k \rho} g_{\lambda (\mu} (x'^{\underline{\beta}},
X_A(y^{\underline{\beta}})) \partial'_{\nu)}
\epsx^\lambda(x'^\gamma, X'_A(y^\gamma)) \nonumber \\& +
\epsx^{\lambda_1 \ldots \lambda_k} (x'^\alpha, X'_A(y^\alpha))
\partial_{\lambda_1 \ldots \lambda_k} g_{\lambda \rho}(x'^\beta,
X_A(y^\beta)) \partial'_\mu
\epsx^\lambda
\partial'_\nu \epsx^\rho(x'^\gamma, X'_A(y^\gamma)) \bigg) \,.
\end{align}
\end{widetext}
Note that according to the Einstein summation convention, repeated
indices are summed over all their possible values; $\epsx^{\lambda_1
\ldots \lambda_k}$ is a short form for the product
$\epsx^{\lambda_1} \ldots \epsx^{\lambda_k}$, and 
$\partial'_{\lambda_1 \ldots \lambda_k}$
stands for $\partial'_{\lambda_1} \ldots \partial'_{\lambda_k}$.
Again, the infinite sum must be regarded as a formal series and does not need
to converge. It has to be truncated at a given order consistently with the
general approximation scheme. There only remains, at this stage, to express the
metric components appearing in the second member of equation
\eqref{eq:dvpt_epsilon} with the help of the new fields $X'_A(y^\alpha)$ rather
than the old $X_A(y^\alpha)$. It can be most easily performed by using the
concept of 
Fr\'echet derivative. Let us consider that
$g_{\mu\nu}(x^\alpha,X_A(y^\alpha))$ is a function of the field variable $X_1$,
on the affine space $\mathbb{R}$, having its domain on an affine space ${\cal
E}_1$ (for instance, the space of smooth bounded real function operating on
$\mathbb{R}^4$, embedded with a norm, say $|~|_{\infty}$). By definition, the
Fr\'echet derivative of 
$g_{\mu\nu}(x^\alpha, X_A(y^\alpha))$ with respect to $X_1$ is, when it exists,
the continuous linear form, on the vector space associated to ${\cal E}_1$,
that satisfies
\begin{multline*}
\mbox{} \! \! \! \! \! g_{\mu\nu}(x^\alpha, X'_1(y^\alpha), X_{A >
  1}(y^\alpha)) - g_{\mu\nu}  
(x^\alpha, X_A(y^\alpha)) \\ = D_{X_1} [g_{\mu\nu}(x^\alpha, X_A(y^\alpha))]
\cdot \delta^*_{\epsx} X_1(y^\beta) + 
o(|\delta^*_{\epsx} X_1(y^\beta)|^2_{\infty})
\end{multline*}
for any infinitesimal variation $\delta^*_{\epsx} X_1(y^\beta)
\equiv X'_1(y^\beta)-X_1(y^\beta) \in {\cal E}_1$. Following usual notations,
we 
have inserted a dot immediately before the variation
$\delta^*_{\epsx} X_1(y^\beta)$ to indicate the action of the form
resulting from the derivation. Denoting by ${\cal L}({\cal E}_1)$ the
vector space of the continuous linear form on ${\cal E}_1$, we have
$D_{X_1} [g_{\mu\nu}(x^\alpha, X_A(y^\alpha))] \in  {\cal L}({\cal E}_1)$.
Now, ${\cal L}({\cal E}_1)$ can be itself embedded with a
structure of affine space as the real number set $\mathbb{R}$, so that we can
define the Fr\'echet derivative of functions $f:{\cal E}_1 \rightarrow {\cal
L}({\cal E}_1)$ in a similar way as before. The bilinear form 
$D_{X_B}[D_{X_1} [g_{\mu\nu}(x^\alpha, X_A(y^\alpha))]]$ is said to be the
second Fr\'echet derivative of $g_{\mu\nu}(x^\alpha, X_A(y^\alpha))$ with
respect to $X_B$ and $X_1$, and
similarly for higher orders. As the derivative of the difference
$g_{\mu\nu}(x^\alpha, X'_1(y^\alpha), X_{A > 1}(y^\alpha)) - g_{\mu\nu}
(x^\alpha, X_A(y^\alpha))$ with respect to $x^\beta$ or $X_B(y^\beta)$
is equal to the difference of the derivatives,
the operator $D_{X_1 \ldots X_{A_k}}$ ($k \in \mathbb{N}$) is evidently
symmetric and commutes with the space-time derivatives
$\partial_{\lambda_1 \ldots \lambda_k}$. We are then in order to achieve our
goal of removing the dependence of the metric on $X_A(y^\beta)$ by expanding
$g'_{\mu\nu}(x^\alpha, X'_A(y^\alpha)-\delta^*_{\epsx}
X_A(y^\alpha))$
around the fields $X'_A(y^\alpha)$ after a generalized Taylor theorem 
\cite{inSchwartz92}. This yields the fundamental relation:
\begin{widetext}
\begin{align} \label{eq:delta_g_NL}
&\delta^*_{\epsx} g_{\mu\nu}  \equiv g'_{\mu\nu} - g_{\mu\nu}
 = \sum_{l=1}^{+\infty} \frac{(-1)^l}{l!}D_{X_{B_1} \ldots
X_{B_l}}[g_{\mu\nu}] . (\delta^*_{\epsx} X_{B_1},
\ldots, \delta^*_{\epsx} X_{B_l}) \nonumber
\\ &+ \sum_{l=0}^{+\infty} \frac{(-1)^l}{l!} \bigg(\epsx^\lambda
\partial_\lambda D_{X_{B_1} \ldots X_{B_l}}[g_{\mu\nu}] \cdot
(\delta^*_{\epsx} X_{B_1}, \ldots, \delta^*_{\epsx}
X_{B_l}) \nonumber
+ 2 D_{X_{B_1} \ldots X_{B_l}}[g_{\lambda
(\mu}] \cdot (\delta^*_{\epsx} X_{B_1}, \ldots,
\delta^*_{\epsx} X_{B_l})
\partial_{\nu)} \epsx^\lambda \bigg)\nonumber \\ &+
\sum_{k,l\ge 0} \! \frac{(-1)^{l}}{k!l!} \bigg(
\frac{\epsx^{\lambda_1 \ldots \lambda_k \lambda \rho}}{(k+1)(k+2)}
\partial_{\lambda_1 
\ldots \lambda_k \lambda \rho} D_{X_{B_1} \ldots X_{B_l}}[g_{\mu\nu}] \cdot
(\delta^*_{\epsx} X_{B_1}, \ldots, \delta^*_{\epsx}
X_{B_l}) \nonumber
\\ & \qquad \qquad \qquad + \frac{2}{k+1} \epsx^{\lambda_1 \ldots
 \lambda_k \rho} \partial_{\lambda_1 \ldots \lambda_k \rho}
D_{X_{B_1} \ldots X_{B_l}}[g_{\lambda (\mu}] \cdot (\delta^*_{\epsx}
X_{B_1}, \ldots, \delta^*_{\epsx} X_{B_l}) \partial_{\nu)}
\epsx^\lambda \nonumber \\ & \qquad \qquad \qquad + \epsx^{\lambda_1 \ldots
\lambda_k} \partial_{\lambda_1 \ldots \lambda_k} D_{X_{B_1} \ldots
X_{B_l}}[g_{\lambda \rho}] \cdot (\delta^*_{\epsx} X_{B_1}, \ldots,
\delta^*_{\epsx} X_{B_l}) \partial_\mu \epsx^\lambda
\partial_\nu \epsx^\rho \bigg) \,, 
\end{align}
\end{widetext}
where 
$(\delta^*_{\epsx} X_{B_1}, \ldots, \delta^*_{\epsx} X_{B_l})$ denotes
the vector to which the $l$-form $D_{X_{B_1} \ldots X_{B_l}} [g_{\mu\nu}]$
applies. It tends toward zero as $c$ goes to infinity since $X'_B$ and $X_B$
coincide when $\epsx^\alpha=0$. As initially required, all
quantities entering the above expression depend implicitly on the same
variables, say $x^\alpha$ and $X_A$. In the grid $x'^\alpha =
x^\alpha - \epsx^\alpha(x'^\beta, X'_B) $, the new components of the metric
are given, at last, by the function $g'_{\mu\nu}(x'^\alpha, X'_A) =
g_{\mu\nu}(x'^\alpha, X'_A) +\delta^*_{\epsx} g_{\mu\nu}(x'^\alpha, X'_A)$. In
the same way, after a further transformation 
$x'^\alpha = x''^\alpha + \varepsilon'^\alpha_{x''}(x''^\beta, X''_B)$, 
the metric components become $g''_{\mu\nu}(x''^\alpha, X''_A) =  g'_{\mu\nu}
(x''^\alpha, X''_A) +\delta^*_{\varepsilon'_{x''}} g'_{\mu\nu}(x''^\alpha,
X''_A)$. This proves the interesting relation:
\begin{align} \label{eq:x_xpp}
x^\mu &= x''^\mu + {\varepsilon'}_{x''}^\mu(x''^\alpha,X''_A) \nonumber \\ +&~
\epsx^\mu(x''^\alpha+{\varepsilon'^\alpha}_{x''}(x''^\beta,X''_B),
X''_A-\delta^*_{\varepsilon'_{x''}} X'_A) \nonumber \\
&= x''^\mu + {\varepsilon'}_{x''}^\mu(x''^\alpha,X''_A) +
\epsx^\mu(x''^\alpha, X''_A) \nonumber \\ +&
\sum_{\genfrac{}{}{0pt}{}{l,k \ge 0}{l+k\ge 1}} \frac{(-1)^l}{k!l!}
{\varepsilon'}_{x''}^{\lambda_1 \ldots \lambda_k} 
\partial''_{\lambda_1 \ldots \lambda_k} D_{X''_{B_1} \ldots
  X''_{B_{l}}} [\epsx^\mu(x''^\alpha,X''_A)] \nonumber \\ & \qquad \qquad
\qquad \qquad \quad \cdot(\delta^*_{\varepsilon'_{x''}}  
X'_{B_1}, \ldots, \delta^*_{\varepsilon'_{x''}} X'_{B_l})  \, .
\end{align}
The preceding transformation between the initial and final coordinates defines
a function $\varepsilon^\mu_{x''}$ by
$x^\mu = x''^\mu + \varepsilon^\mu_{x''}$. Equation
\eqref{eq:x_xpp} shows that this function is not the mere superposition of
$\epsx^\mu$ and $\varepsilon'^\mu_{x''}$ but also involves non-linear
contributions. We already know that the metric components $g'_{\mu\nu}$
obtained after the first coordinate change are given by 
$g'_{\mu\nu} = g_{\mu\nu} + \delta^*_{\epsx} g_{\mu\nu}$. 
The ten functions $g''_{\mu\nu}$ can be similarly deduced from $g'_{\mu\nu}$:
$g''_{\mu\nu} = g'_{\mu\nu} + \delta^*_{\varepsilon'_{x''}} g'_{\mu\nu}$. On
the other hand, as the grid $\{x''^\mu\}$ is related to the old coordinate
system $\{x^\mu\}$ by $x^\alpha = x''^\alpha + \varepsilon_{x''}^\alpha$, 
we have
$g''_{\mu\nu} = g_{\mu\nu} + \delta^*_{\varepsilon_{x''}} g_{\mu\nu}$
(which can be checked explicitly). Finally, the quantities 
$g'''_{\mu\nu} = g_{\mu\nu} + \delta^*_{\epsx^\mu + \varepsilon'^\mu_{x''}}
g_{\mu\nu}$   
represent the components of the metric in the coordinates system 
$\{x{'''}^\alpha\}$ such that 
$x^\alpha = x{'''}^\alpha +\epsx^\alpha(x{'''}^\beta, X{'''}_{\! \! \! \! \! \!
  A \,}(y^\beta)) +  
\varepsilon'^\alpha_{x''}(x{'''}^\beta, X{'''}_{\! \! \! \! \! \! A
  \,}(y^\beta))$. They refer to the same 
gravitational field as $g'_{\mu\nu}$, or $g''_{\mu\nu}$.
In all what precedes, the new 
coordinates are defined by implicit equations for given $\epsx^\alpha$ and
$\varepsilon'^\alpha_{x''}$, e.g.
$x'^\alpha = x^\alpha - \epsx^\alpha (x'^\beta,X'_A(y^\beta))$. However, they
can be formally inverted by applying recursively the Taylor formula and
rearranging the summation,
\begin{widetext}
\begin{multline*}
x'^\alpha = x^\alpha - \sum_{n,m \ge 0}^{+\infty} 
\frac{(-1)^n}{m!} \sum_{s=1}^{+\infty} \sum_{\genfrac{}{}{0pt}{}{k_1 +
\ldots + k_s = n}{k_i \ge 1}} \frac{1}{k_1 ! \ldots k_s !} 
D[\varepsilon^{\lambda_1^{(n)} 
\ldots \lambda_{k_n}^{(n)}} \partial_{\lambda_1^{(n)} \ldots
\lambda_{k_n}^{(n)}} \times\\ \times \varepsilon^{\lambda_1^{(n-1)}
\ldots \lambda_{k_{n-1}}^{(n-1)}} 
\ldots \partial_{\lambda_1^{(2)} \ldots \lambda_{k_2}^{(2)}}
\varepsilon^{\lambda_1^{(1)} \ldots \lambda_{k_1}^{(1)}}
\partial_{\lambda_1^{(1)} \ldots \lambda_{k_1}^{(1)}} \varepsilon^\alpha]\cdot
(\delta^* X_{B_1}, \ldots, \delta^* X_{B_m}) \, .
\end{multline*}

In most of applications, the post-Newtonian metric components show the
structure of a sum of products of elementary factors, each of them being itself
a $n$-tuple spatial integral of a given function $f(\mathbf{y}, \mathbf{z}_1,
\ldots, \mathbf{z}_{n-1}, t, X_A(\mathbf{y},c t))$ of the coordinates and the
fields. With the help of the Leibniz
rule, the Fr\'echet derivative $D_{X_A} [g_{\mu\nu}(x^\alpha, X_B(y^\alpha))]$
can be written as a sum of terms matching the pattern:
\[F(x^\alpha, X_C(y^\alpha)) D_{X_A}\bigg[ \integ d^3\mathbf{y}_1 \ldots
d^3{\bf y}_n~ f({\bf y_1}, {\bf y_2}, \ldots, {\bf y_n}, t,X_D({\bf y_1},c
t))\bigg]\cdot \delta^*_{\epsx} 
X_B \, , \] 
where $F(x^\alpha, X_C(y^\alpha))$ is a function of $x^\alpha$ and a
functional of $X_C(y^\alpha))$. After the definition of $D_{X_A}$, the latter
expression also reads:
\[\integ d^3\mathbf{y}_1~\delta^*_{\epsx} X_A \integ d^3\mathbf{y}_2 \ldots
d^3\mathbf{y}_n~ F(x^\alpha, X_C(y^\alpha)) \partial_{X_A}f({\bf y_1}, {\bf
  y_2}, \ldots, {\bf y_n}, t,X_D({\bf
y_1},c t))\, . \] 
From what we conclude that $D_{X_A} [g_{\mu\nu}(x^\alpha,  X_B(y^\alpha))].
\delta^*_{\epsx} X_A$ 
admits an integral representation of the type 
$\integ d^3\mathbf{y} ~\delta^*_{\epsx} X_A (\delta g_{\mu\nu}/\delta
X_A)$. The so-defined quantity $\delta g_{\mu\nu}/\delta X_A$ is known as the
first functional derivative of $g_{\mu\nu}(x^\alpha,  X_B(y^\alpha))$ with
respect to $X_A$. It is also possible to introduce the second
functional derivative $\delta^2 g_{\mu\nu}/$ $\delta X_A \delta X_B$. By
definition,
\begin{equation*}
 \integ d^3\mathbf{y}_1 d^3\mathbf{y}_2~ \delta^*_{\epsx} X_B(\mathbf{y}_1,ct)
\delta^*_{\epsx}  X_C(\mathbf{y}_2,ct)  \frac{\delta^2
g_{\mu\nu}}{\delta^*_{\epsx} X_B \delta^*_{\epsx} X_C}
\equiv D_{X_B X_C} [g_{\mu\nu}(x^\alpha,
X_A(y^\alpha))] \cdot (\delta^*_{\epsx} 
X_B(y^\beta), \delta^*_{\epsx} X_C(y^\gamma)) \, .
\end{equation*}
\end{widetext}

The case where the non-linear terms occurring in \eqref{eq:delta_g_NL} do not
contribute to $\delta^*_{\epsx} g_{\mu\nu}$ (or at least to the
order of $g_{\mu\nu}$ we are interested in) is of particular importance. This
happens when all products involving two factors equal to
$\delta^*_{\epsx} X_B$ or $\epsx$ as well as a space or time derivative of the
metric 
$\partial g$ can be neglected. Hence, the $n$ PN transformation law
is linear whenever $\cO(1/c^{n+1}) =\cO(\epsx^2 \partial g) +
\cO(\epsx \delta^*_{\epsx} X \partial g) + \cO(( \delta^*_{\epsx} X)^2\partial
g)$, and equation
\eqref{eq:delta_g_NL} reduces then to:
\begin{multline} \label{eq:delta_g_L}
\delta^*_{\epsx} g_{\mu\nu} = -
D_{X_B}[g_{\mu\nu}] \cdot \delta^*_{\epsx} X_B \\ + \epsx^\lambda 
\partial_\lambda g_{\mu\nu} + 2 g_{\lambda (\mu} \partial_{\nu)}
\epsx^\lambda \, .
\end{multline}

As a check, let us specify relation \eqref{eq:delta_g_L} to the important case
where the matter variables are (\emph{i}) the baryonic density of the fluid 
$\rhoc = \sqrt{-g} \rho u^0$ (with $g = \det g_{\mu\nu}$), (\emph{ii}) the 
fluid coordinate 3-velocity field $v^i = c u^i/u^0$, and (\emph{iii}) the
entropy per unit mass $s$, i.e. 
$g_{\mu\nu} = g_{\mu\nu}(x^\alpha,\rhoc,v^p,s)$. 

By using the transformation properties of $\rho$ (scalar), $u^\mu$ (vector)
and $\sqrt{-g}$ (density), we find at the linear order after a straightforward
calculation:
\begin{align*} 
\delta^* \rhoc =&~ \partial_i (\rhoc \epsx^i) - \frac{v^i}{c} \partial_i
\epsx^0 \rhoc
+ \frac{\epsx^0}{c} \partial_t \rhoc \\ =&~ \partial_i (\rhoc \epsx^i - \rhoc
\frac{v^i}{c} \epsx^0)\, ,
\\
\delta^* v^i =&~ \frac{1}{c} \partial_t (\epsx^0 v^i) + \epsx^j
\partial_j v^i - v^j \partial_j \epsx^i \\ & 
-\partial_t \epsx^i + \frac{v^i v^j}{c} \partial_j \epsx^0 \\ =&~
\frac{d}{dt} (-\epsx^i + \frac{v^i}{c} \epsx^0) - (- \epsx^j +
\frac{\epsx^0}{c} v^j) \partial_j v^i\, , \\
\delta^* s =&~ \frac{\epsx^0}{c} \partial_t s + \epsx^i \partial_i s\, , 
\end{align*}
having made use of the continuity equation to obtain the second expression of
$\delta^* \rhoc$ (see section \ref{sec:reaction}).
The variation $\delta^* v^i$ is put into a rather compact form
by resorting to the total derivative $d/dt = \partial_t + v^i
\partial_i$. By virtue of equation \eqref{eq:delta_g_L}, the new components of
the metric read 
\begin{multline}
g'_{\mu\nu} = g_{\mu\nu}-D_{\rhoc}[g_{\mu\nu}]\cdot \delta^* \rhoc -
D_{v^q}[g_{\mu\nu}].\delta^* v^q - D_{s}[g_{\mu\nu}]\cdot \delta^* s
\\ + \epsx^\lambda \partial_\lambda
g_{\mu\nu} + 2 g_{\lambda(\mu} \partial_{\nu)} \epsx^\lambda\, .
\end{multline}
If the source is made of one point particle of mass $m$, then
$\rhoc(\mathbf{y}, t) = m \delta(\mathbf{y} - \mathbf{z}(t))$, 
and  the metric can be viewed as a functional of $\mathbf{z}$ and
$\mathbf{v}$,
\[
g_{\mu\nu}(x^\alpha, \mathbf{z}(t), \mathbf{v}(t)) = \frac{1}{m} \integ d^3
\mathbf{y}~ g_{\mu\nu} (x^\alpha, \mathbf{y}, \mathbf{v}(\mathbf{y},t))~
\rhoc(\mathbf{y},t) \, . 
\]
The contribution of the functional derivatives is
\begin{align*}
&- m (D_{\rhoc}[g_{\mu\nu}]\cdot
\delta^* \rhoc + D_{v^q}[g_{\mu\nu}]\cdot \delta^* v^q)
\\ & = - \integ d^3 \mathbf{y}~\{g_{\mu\nu}(x^\alpha, \mathbf{y},
\mathbf{v}(\mathbf{y},t)) 
 \delta^* \rhoc \\ & \qquad + D_{v_q} [g_{\mu\nu}(x^\alpha, {\bf
   y}, \mathbf{v}(\mathbf{y},t))] \cdot \delta^* v^q \rhoc\} \\ &=
\integ d^3 \mathbf{y}~\{ (D_{y^i}[g_{\mu\nu}(x^\alpha, \mathbf{y},
\mathbf{v}(\mathbf{y},t))] \\ & \qquad + \partial_i v^j D_{v^j}
[g_{\mu\nu}(x^\alpha, \mathbf{y}, \mathbf{v}(\mathbf{y},t))] )\}  \cdot
\Big(\rhoc \epsx^i - \rhoc \frac{v^i}{c}  
\epsx^0\Big) \\ & \qquad  - \rhoc D_{v_i} [g_{\mu\nu}(x^\alpha, {\bf
y}, \mathbf{v}(\mathbf{y},t))]\} \cdot \bigg(\frac{d}{dt} \Big(-\epsx^i +
\frac{v^i}{c} \epsx^0\Big) \\ & \qquad \qquad \qquad \qquad \qquad \qquad
\qquad - \Big( - \epsx^j + \frac{\epsx^0}{c} v^j\Big) \partial_j v^i\bigg)\\
& = - m \integ d^3 \mathbf{y} ~\{
(-\epsx^i + \frac{v^i}{c} \epsx^0) D_{y^i}[g_{\mu\nu}(x^\alpha,
\mathbf{y}, \mathbf{v}(\mathbf{y},t))] \\ & \qquad + \frac{d}{dt}
(-\epsx^i + \frac{v^i}{c} \epsx^0) \\ & \qquad \quad \times
D_{v^i}[g_{\mu\nu}(x^\alpha, \mathbf{y}, {\bf 
v}(\mathbf{y},t))]  \} \delta(\mathbf{y}-\mathbf{z}(t))\, , 
\end{align*}
which shows that
\begin{multline*} D_{\rhoc}[g_{\mu\nu}] \cdot
\delta^* \rhoc + D_{v^q}[g_{\mu\nu}]\cdot \delta^* v^q \\ = 
D_{z^i}[g_{\mu\nu}] \cdot \delta^* z^i +D_{v^i}[g_{\mu\nu}] \cdot 
\frac{d}{dt}\delta^* z^i \, ,
\end{multline*}
where $\delta^* z^i = -\epsx^i(\mathbf{z}(t),t) + v^i(t)
\epsx^0(\mathbf{z}(t),t)/c$. This is in perfect agreement with the 
linearized boost transformation formulas \cite{DS85}: 
$c t' = c t - \epsx^0(\mathbf{z}(t),t)$;
$z'^i(t') = z^i(t) -\epsx^i(\mathbf{z}(t),t)$.

\section{Elimination of highest time derivatives in the metric of a
general perfect fluid} \label{sec:gauge}
An analytic expression for that part of the metric that contributes to the
$3.5$ PN reaction force arising 
in an isolated system due to gravitational wave emission is available in a
Burke-Thorne-like gauge which also fulfills the harmonicity
conditions at 1 PN order \cite{B97}, for general smooth, spatially compact
distributions of matter.  However, it is not suited for
numerical calculations for it depends on seventh order time derivatives of
certain multipole moments. A possible cure consists in performing a change of
coordinates such that terms of $\delta^*_{\epsx} g_{\mu\nu}$ cancel the
highest time derivatives in the relevant part of $g_{\mu\nu}$
(the functional dependence will not be specified anymore henceforth). 
In order to be able to apply the formula \eqref{eq:delta_g_L}, we first have
to make choice of the fields $X_A$. It is convenient to take the mass density
$\sigma = (T^{00}+T^{ii})/c^2$, the current density $\sigma_i = T^{0i}/c$, and
the stress $\sigma_{ij}= T^{ij}$, following paper \cite{B97}. All three
quantities have compact spatial support and are Newtonian at the
leading order for weakly stressed systems. Moreover, they
lead to a simple expression for the metric, based on a parametrization by
means of certain potentials of which they are the sources. At the
1 PN order, these are the Poisson integrals:
\begin{align}
 U({\mathbf{x},t}) &= G  \integ
 \frac{d^3\mathbf{y}}{|\mathbf{x}-\mathbf{y}|}~\sigma({{\bf y},t}) \,
 ,\nonumber \\  U_i({\mathbf{x},t}) &= G \integ
 \frac{d^3\mathbf{y}}{|\mathbf{x}-\mathbf{y}|}~ \sigma_i({\mathbf{y},t}) \, ,
\end{align}
as well as the super potential 
$\chi({\mathbf{x},t}) = G \integ d^3\mathbf{y} ~|\mathbf{x}-\mathbf{y}|
\sigma({\bf y},t)$. They represent the Poisson 
solutions of the equations $\Delta U \equiv \partial_i \partial_i U = -4\pi G
\sigma$, $\Delta U_i = -4\pi G \sigma_i$, $\Delta \chi = 2 U$ respectively.
We shall write $U = \Delta^{-1} (-4\pi G \sigma)$ or 
$\chi = \Delta^{-2} (-8\pi G \sigma)$. It is worth noticing that $U$
reduces to the Newtonian potential when $c$ goes to infinity. The
terms of the metric that contribute to the $1/c^5$
and $1/c^7$ part of the equations of motion take also a particularly simple
form. They indeed depend on only two ``reaction''
potentials $U^{\rm reac}$ and $U_i^{\rm reac}$ involving the mass
quadrupole  $I_{ij}(t)$, the mass octupole $I_{ijk}(t)$, and the current
quadrupole $J_{ij}(t)$ as defined in the Blanchet-Damour formalism (see
\cite{BH97} and references therein):
\begin{subequations}
\begin{align} \label{eq:Ureac}
U^{\rm reac}(\mathbf{x},t) =& - \frac{G}{5c^5} x^{ij} I^{(5)}_{ij}(t) +
\frac{G}{c^7} \bigg[ \frac{1}{189} x^{ijk} I^{(7)}_{ijk}(t)  \nonumber  
\\ & - \frac{1}{70} x^{kk} x^{ij} I^{(7)}_{ij}(t) \bigg] + \cO
\left(\frac{1}{c^8}\right) \, ,\\
U^{\rm reac}_i(\mathbf{x},t) =&~ \frac{G}{c^5} \bigg[\frac{1}{21} \hat{x}^{ijk}
I^{(6)}_{jk}(t) \nonumber \\ & - \frac{4}{45} \epsilon_{ijk} x^{jm}
J^{(5)}_{km}(t) \bigg]+ \cO\left( \frac{1}{c^6} \right) \, ,
\end{align}
\end{subequations}
the symbol $\hat{x}^{ijk\ldots}$, or equivalently $x^{<ijk\ldots>}$, being a
short form for the symmetric
trace-free (STF) part of $x^{ijk\ldots} \equiv x^i x^j x^k \ldots $, 
and $I^{(n)}_{ij}(t)$, an alternative notation for the $n$th time derivative 
$d^n I_{ij}(t)/dt^n$. These
potentials differ from $V^{\rm reac}$ and $V_i^{\rm reac}$ used in \cite{BH97}
respectively by terms of order $1/c^8$ and $1/c^6$ which do 
not play any role at our level \cite{B97}. We give for completeness the 
multipole moments entering $U^{\rm reac}$ and $U_i^{\rm reac}$ as functionals
of $\sigma$ and $\sigma_i$: 
\begin{subequations} \label{eq:moments}
\begin{align}
I_{ij} &= \integ d^3\mathbf{y} \left(\hat{y}^{ij} \sigma+ \frac{1}{14c^2}
  y^{kk} \hat{y}^{ij} \partial_t^2 \sigma - \frac{20}{21c^2} \hat{y}^{ijk}
  \partial_t \sigma_k \right) \, , \\
I_{ijk} &= \integ d^3\mathbf{y}~\hat{y}^{ijk} \sigma + \cO\left(\frac{1}{c^2} 
\right) \, , \\
J_{ij} &= \integ d^3\mathbf{y}~ \epsilon_{km<i} \hat{y}^{j>k} \sigma_m \, ,
\end{align}
\end{subequations}
with $\epsilon_{kmi}$ holding for the Euclidean Levi-Civita tensor (in three
dimensions). Having the potentials $U$, $U_i$, $\chi$,
$U^{\rm reac}$, and $U_i^{\rm reac}$ to our disposal, we can write the
metric components as
\begin{subequations} \label{eq:metric}
\begin{align} 
\mbox{} \! \! \! g_{00} =& -1 + \frac{2}{c^2} (U+U^{\rm reac})
+\frac{1}{c^4} \left[\partial_t^2 \chi -2 U^2 - 4 U U^{\rm reac}
\right] \nonumber \\ & + (g_{00})_{(6+8)} + \cO
\left(\frac{1}{c^{10}} \right) \, ,  \\ 
\mbox{} \! \! \! g_{0i} =& -\frac{4}{c^3} (U_i+U_i^{\rm reac}) +
(g_{0i})_{(5+7)}+ \cO\left(\frac{1}{c^9} 
\right) \, , \\
\mbox{} \! \! \! g_{ij} =&~ \delta_{ij} \left[1+ \frac{2}{c^2} (U+U^{\rm
    reac})\right]  + 
(g_{ij})_{(4+6)} + 
\cO\left(\frac{1}{c^8} \right) \, ,
\end{align}
\end{subequations}
where the indices between brackets indicate the order of the
term they refer to, so that ${g_{00}}_{(6+8)}$ denotes the full
set of $1/c^6$ and $1/c^8$ terms entering $g_{00}$.

It should be stressed that the $1/c^2$ order in
the corrections to the flat metric entering the right-hand side of equation
\eqref{eq:metric} is \emph{not}
the post-Newtonian order, the latter being rather defined as
the level of contributions \emph{to the equations of
motion}. Since $(g_{00})_{(2)}$ is responsible for the Newtonian force,
the terms $(g_{00})_{(4)}$, $(g_{0i})_{(3)}$, and $(g_{ij})_{(2)}$ that
come just after in the power expansion are evidently post-Newtonian. 
When the matter variables $X_A$ are chosen in a way they have a non-zero
Newtonian limit $X_{A(0)} = X_A + \cO(1/c^2)$ and derive from the trajectory
of the fluid in the configuration space independently from the metric, 
the $1/c^5$ and $1/c^7$ terms of the equations of motion
identify with the reaction force in the post-Newtonian scheme. 
Other set of variables of the type $X_A = X_{A(0)} + \cO(1/c^2)$ with 
$X_{A(0)} \neq 0$ may be
equivalently used provided all expansions are performed consistently including
the Newtonian, 1 PN, 2.5 PN, and 3.5 PN terms. 
The coupling
between space-time curvature and radiation being a 4 PN effect, the even orders
are purely conservative. It is thus convenient to split
$g_{\mu\nu}$ into an odd $(g_{\mu\nu})_{\rm odd}$ and an even part 
$(g_{\mu\nu})_{\rm even}$, named after the parity of the terms
they generate in the equations of motion expressed with the help of the
variables that have been adopted.
As an example, $(g_{00})_{(7)}$, $(g_{0i})_{(6)}$ or $(g_{ij})_{(5)}$ will
belong to the odd part, while $(g_{00})_{(2)}$,
$(g_{0i})_{(3)}$ or $(g_{ij})_{(2)}$ will enter the even one.
In our problem, the odd (resp. even) part includes the $2.5$ and $3.5$ PN 
(resp.
Newtonian and post-Newtonian) corrections. The even part corrections that are
beyond the 1 PN approximation do not play any role and can be let unspecified.

We propose now to conduct a coordinate transformation intended 
to reduce, as far as possible, the order of time derivatives appearing in the
original metric components. This can be achieved by eliminating the
potentials $U^{\rm reac}_{(7)}$ and $U^{\rm reac}_{i(5)}$ containing the
highest orders of 
derivation. In addition, we require that the new gauge identifies with
the standard post-Newtonian one,  
$\partial_i g'_{0i} -\frac{1}{2} \partial_0 g'_{ii} = \cO(1/c^5)$, 
$\partial_j g'_{ij} - \frac{1}{2} \partial_i (g'_{jj} - g'_{00}) = \cO(1/c^4)$,
which is itself equivalent to the Arnowitt-Deser-Misner
(ADM) gauge \cite{ADM62} up to the 1 PN order.
As the original 1 PN gravitational field \eqref{eq:metric} satisfies the
harmonicity conditions
in the near zone (i.e. the region of space including the system in
which the post-Newtonian approximation is valid) modulo 1.5 PN corrections,
the functions $\epsx^\mu$ corresponds \emph{at the leading order} to
the difference between harmonic and ADM coordinates. After paper \cite{DS85},
we have $\epsx^0 = \cO(1/c^3)$ and $\epsx^i=\cO(1/c^4)$,
which shows that non-linear terms of the kind 
$\epsx^{\lambda_1\lambda_2\ldots} \partial_{\lambda_1 \lambda_2\ldots}
D_{\ldots}[g_{\mu\nu}]\ldots = \cO(1/c^{10})$ drop out in 
equation \eqref{eq:delta_g_NL}. The only non-linearities that may remain are
those generated by the functional variations $\delta^*_{\epsx} \sigma$, 
$\delta^*_{\epsx} \sigma_i$, and
$\delta^*_{\epsx} \sigma_{ij}$. To know whether they can be
neglected or not, we have to determine the post-Newtonian order of the three
latter quantities. The linear approximation is certainly
sufficient for this purpose. We compute
$\sigma(y'^\alpha,g'_{\alpha\beta}(z^\gamma,X'_B))$,
$\sigma_i(y'^\alpha,g'_{\alpha\beta}(z^\gamma,X'_B))$, and
$\sigma_{ij}(y'^\alpha,g'_{\alpha\beta}(z^\gamma,X'_B))$, using the gauge 
transformation law
$ T^{\mu\nu}(y'^\alpha,g'_{\alpha\beta}(z^\gamma,X'_B)) =
T^{\mu\nu}(y^\alpha,g_{\alpha\beta}(z^\gamma,X_B))  - 2 \partial_\lambda
\epsx^{(\mu} T^{\nu)\lambda}(y^\alpha,g_{\alpha\beta}(z^\gamma,X_B)) +
\cO(\varepsilon^2)$
for the stress-energy tensor. After straightforward calculations, we get
\begin{subequations}
\begin{align}
\delta^*_{\epsx} \sigma =&~ \epsx^k \partial_k \sigma -
\frac{2}{c} \sigma \partial_t \epsx^0 + \frac{1}{c} \epsx^0 \partial_t
\sigma - \frac{2}{c} \sigma_i \partial_i \epsx^0 \nonumber \\ & -
\frac{2}{c^2} \sigma_i 
\partial_t  \epsx^i - \frac{2}{c^2} \sigma_{ik} \partial_k \epsx^i
\nonumber \\ &  + \frac{2}{c^3} \sigma_{kk} \partial_t \epsx^0
+\cO(\varepsilon^2) + \cO((\delta^*_{\epsx} 
g)^2) \, , \\ \delta^*_{\epsx} \sigma_i =&~ -\sigma \partial_t
\epsx^i - \sigma_j \partial_j \epsx^i + \epsx^j \partial_j \sigma_i -
\frac{1}{c} \sigma_i \partial_t \epsx^0 \nonumber \\ & + \frac{1}{c} \epsx^0
\partial_t \sigma_i - \frac{1}{c} \sigma_{ij} \partial_j \epsx^0 \nonumber \\
& + \frac{1}{c^2} \sigma_{jj} \partial_t \epsx^i +
\cO(\varepsilon^2) +  \cO((\delta^*_{\epsx} g)^2) \, , \\
\delta^*_{\epsx} \sigma_{ij} =&~ - 2 \sigma_{(i} \partial_t
\epsx^{j)} - 2  \partial_k \epsx^{(i} \sigma_{j)k}+ \epsx^k
\partial_k \sigma_{ij} \nonumber \\ & + \frac{1}{c} \epsx^0 \partial_t
\sigma_{ij} + \cO(\varepsilon^2) + \cO((\delta^*_{\epsx}
g)^2) \, ,
\end{align}
\end{subequations}
which means notably that 
$\delta^*_{\epsx} {X_A} = \cO(1/c^4) = \cO (\epsx^i, \epsx^0/c )$.
Consequently, there cannot be nonlinear contributions in 
$\delta^*_{\epsx} g_{\mu\nu}$, and equation \eqref{eq:delta_g_L} yields
\begin{subequations} \label{eq:deltag}
\begin{align}
(\delta^*_{\epsx} g_{00})_{\rm odd} =&~ - \frac{2}{c} \partial_t
{\epsx^0}_{(6+8)} + \frac{4}{c^3} U \partial_t {\epsx^0}_{(6)} \nonumber \\ &
+ \frac{2}{c} 
(g_{0i})_{(3)} \partial_t {\epsx^i}_{(5)} + \frac{2}{c^3} {\epsx^0}_{(6)}
\partial_t U  \nonumber \\ & + {\epsx^i}_{(5)} \partial_i (g_{00})_{(4)} + 
\frac{2}{c^2} {\epsx^i}_{(5+7)} \partial_i U \nonumber \\ & - \frac{2}{c^2}
\integ  d^3\mathbf{y}~\frac{\delta U}{\delta X_A} (\delta^*_{\epsx}
X_A)_{(5+7)} \nonumber \\ & - \integ d^3\mathbf{y}~\frac{\delta
(g_{00})_{(4)}}{\delta X_A} (\delta^*_{\epsx} X_A)_{(5)} \, ,
\label{eq:deltag00} 
\\ (\delta^*_{\epsx} g_{0i})_{\rm odd}
=&~ - \partial_i {\epsx^0}_{(6+8)} + \frac{1}{c} \partial_t {\epsx^i}_{(5+7)}
\nonumber \\ & + \frac{2}{c^2} U \partial_i {\epsx^0}_{(6)} + \frac{2}{c^3} U
\partial_t {\epsx^i}_{(5)} \nonumber \\ & +  (g_{0j})_{(3)} \partial_i
{\epsx^j}_{(5)} + {\epsx^j}_{(5)} \partial_j (g_{(0i)})_{(3)} 
\nonumber \\ & - \integ 
d^3\mathbf{y}~\frac{\delta (g_{0i})_{(3)}}{\delta X_A} (\delta^*_{\epsx}
X_A)_{(5)}
\, , \label{eq:deltag0i}
\\ (\delta^*_{\epsx} g_{ij})_{\rm odd} =&~ 2
\partial_{(i} {\epsx^{j)}}_{(5+7)} + \frac{4}{c^2} U \partial_{(i}
{\epsx^{j)}}_{(5)} \nonumber \\ & + \frac{2}{c^2} {\epsx^k}_{(5)}  \partial_k
U \delta^{ij} \nonumber \\ & - \frac{2}{c^2} \delta^{ij} \integ d^3{\bf
  y}~\frac{\delta U}{\delta 
X_A}  (\delta^*_{\epsx} X_A)_{(5)} \, . \label{eq:deltagij}
\end{align}
\end{subequations}

On the other hand, relations \eqref{eq:moments} and \eqref{eq:metric} tell us
that the time derivatives of highest order in $g'_{\mu\nu}$, namely
$\partial_t^6$ and $\partial_t^7$, come (\emph{i}) from the first correction
$U^{\rm reac}_{(7)}$ of the reaction potential $U^{\rm reac}$ that
parametrizes the $3.5$ PN part of $g'_{00}$, and (\emph{ii}) from the leading
approximation of the reaction potential that contributes to the $3.5$ PN part
of $g'_{0i}$, $U_i^{\rm reac} = U_{i(5)}^{\rm reac}$. We shall restrict
ourselves to the case where the new metric can be identified with a function
of certain elementary potentials $P_1, P_2, \ldots$ These potentials are
supposed to be spatial integrals whose sources are \emph{a priori} products of
(\emph{i}) densities such as $\sigma$, $\sigma_i$, $\sigma_{ij}$, or their
derivatives evaluated at the location of the integration variable $\mathbf{y}$
at time $t$, (\emph{ii}) some factors depending on the field point
$\mathbf{x}$, and possibly (\emph{iii}) instantaneous potentials of the same
kind. Typical examples are $U$, $U_i$, or $U^{\rm reac}_{(5)}$. That the
latter quantity is indeed of the required type, this can be seen by letting
the factor $x^{ij}$ and the time derivatives go under the integration symbol
in the first term of relation \eqref{eq:Ureac}. Following this hypothesis, we
shall search for $\epsx^\mu$ as a function of $U$, $U_i$, or some other
potentials $P_1, P_2 \ldots$ Moreover, our choice must be such that at least
$U_{(7)}^{\rm reac}$ simplifies with terms coming from the gauge
transformation \eqref{eq:deltag}.

To be more explicit, we come
back to equations \eqref{eq:deltag00}. Together with equations
\eqref{eq:metric}, they imply
\begin{subequations}\label{eq:gnew}
\begin{align} 
(g'_{00})_{(7+9)} =&~  (h_{00})_{(9)} - \frac{2}{c} \partial_t
{\epsx^0}_{(6+8)} +
\frac{2}{c^2} {\epsx^i}_{(5+7)} \partial_i U \nonumber \\ 
& - \frac{2}{c^2} \integ
d^3\mathbf{y}~\frac{\delta U}{\delta \sigma} (\delta^*_{\epsx}
\sigma)_{(5+7)} + \frac{2}{c^2} U^{\rm reac}_{(5+7)} \, , \label{eq:g00new}
\\ (g'_{0i})_{(6+8)} =&~ (h_{0i})_{(8)} - \partial_i {\epsx^0}_{(6+8)}
\nonumber \\ & + \frac{1}{c} \partial_t {\epsx^i}_{(5+7)} - \frac{4}{c^3}
U_{i(5)}^{\rm reac} 
\, , \label{eq:g0inew} \\
(g'_{ij})_{(5+7)} =&~  (h_{ij})_{(7)} + 2 \partial_{(i} {\epsx^{j)}}_{(5+7)}
+ \frac{2}{c^2} \delta^{ij} U^{\rm reac}_{(5)} \, . \label{eq:gijnew}
\end{align}
\end{subequations}
For simplicity, the $3.5$ PN terms that are
functionals of ${\epsx^0}_{(6)}(P_1,P_2,\ldots)$ or
${\epsx^i}_{(5)}(P_1,$ $P_2,\ldots)$ and functions of potentials
involving second order time derivatives, at most, were gathered in the matrix
$h_{\mu\nu}$. At the leading order, equation \eqref{eq:g00new} reduces to
\begin{multline*}
 (g'_{00})_{(7)} = - \frac{2}{c} \partial_t {\epsx^0}_{(6)} + \frac{2}{c^2} 
{\epsx^i}_{(5)} \partial_i U \\ - \frac{2}{c^2} \integ
d^3\mathbf{y}~\frac{\delta U}{\delta \sigma} (\delta^*_{\epsx} 
\sigma)_{(5)} + \frac{2}{c^2} U^{\rm reac}_{(5)} \, .
\end{multline*}
As long as $U^{\rm reac}_{(5)}$ appears explicitly in the
gravitational field, we shall have to evaluate the fifth order time
derivative $I_{ij}^{(5)}(t)$ in numerical simulations. We can avoid this
by imposing that the reaction potential $U^{\rm reac}_{(5)}$ cancels the
contribution to $(g'_{00})_{(7)}$ that is linear in ${\epsx^0}_{(6)}$ or
${\epsx^i}_{(5)}$, which amounts to demanding that
${\epsx^{0}}_{(6)}$ (resp. ${\epsx^i}_{(5)}$) decomposes into 
terms containing time derivatives of fourth order at most, plus terms 
${\epsx^0}^{\rm (part)}_{(6)}$ (resp. 
${\epsx^i}^{\rm (part~I)}_{(5)}$ or ${\epsx^i}^{\rm (part~II)}_{(5)}$)
satisfying
\begin{multline*}
- \frac{2}{c} \partial_t {\epsx^0}^{\rm (part)}_{(6)} + \frac{2}{c^2} 
{\epsx^i}^{\rm (part~I)}_{(5)} \partial_i U \\ - \frac{2}{c^2} \integ
d^3\mathbf{y}~\frac{\delta U}{\delta \sigma} {\epsx^k}^{\rm (part~II)}_{(5)}
\partial_k \sigma + \frac{2}{c^2} U^{\rm reac}_{(5)} = 0 \, .
\end{multline*}
While either $2 {\epsx^i}^{\rm (part~I)}_{(5)} \partial_i U/c^2$
or $- 2 \integ d^3\mathbf{y}~(\delta U/\delta \sigma)$
${\epsx^k}^{\rm (part~II)}_{(5)} \partial_k \sigma/c^2$ contribute to the
cancellation of $2 U^{\rm reac}_{5}/c^2$, the quantities
${\epsx^i}_{(5)}^{\rm part (I)}$ or ${\epsx^i}_{(5)}^{\rm part (II)}$ are, by
contrast, explicit functions of $U^{\rm reac}_{(5)}$. If ${\epsx^i}_{(5)}^{\rm
part (I)} \neq 0$ or ${\epsx^i}_{(5)}^{\rm part (II)} \neq 0$, the
fifth derivative $I_{ij}^{(5)}(t)$ still occurs in the metric by virtue of the
relation $g'_{ij} = 2 \partial_{(i} {\epsx^{j)}}_{(5)} + \cO(1/c^7)$. 
This eventuality being rejected, $2 U^{\rm reac}_{(5)}/c^2$ must simplify with 
$- 2\partial_t {\epsx^{0}}_{(6)}^{\rm (part)}/c$ exclusively, i.e.
\[- \frac{2}{c} \partial_t {\epsx^0}^{\rm (part)}_{(6)} + 
\frac{2}{c^2} U^{\rm reac}_{(5)} = 0 \, . \]
Therefore, we may construct ${\epsx^0}_{(6)}$ as the sum of $U^{{\rm
reac}(-1)}_{(5)}/c = - G x^{ij} I^{(4)}_{ij}(t)/(5 c^6)$
(antiderivative of $U^{\rm reac}_{(5)}/c$ that vanishes when $I_{ij}(t)=0$),
plus some arbitrary function ${\epsx^0}_{(6)}^{\rm (arb)}$ that does
not involve derivatives of order higher than four:
\begin{equation} \label{eq:eps06_part_arb}
{\epsx^0}_{(6)} = \frac{1}{c} U^{{\rm reac}(-1)}_{(5)} +
{\epsx^0}_{(6)}^{\rm (arb)}  \, .
\end{equation}
As a matter of fact, the choice of ${\epsx^0}_{(6)}^{\rm (arb)}$ together with
that of ${\epsx^i}_{(5)}$ entirely determines the maximal order 
$n_d( (g'_{00})_{(7)} )$ of time derivations
in $(g'_{00})_{(7)}$. Next, the expression for the odd $0i$
components of the new metric \eqref{eq:g0inew} at order $2.5$ PN,
\[ (g'_{0i})_{(6)} = - \partial_i {\epsx^0}_{(6)} +
\frac{1}{c} \partial_t {\epsx^i}_{(5)} \, , \]
shows that $n_d ( (g'_{0i})_{(6)} ) = n_d ( \partial_i
{\epsx^0}_{(6)} ) = 4$, \emph{unless} $-\partial_i {\epsx^0}_{(6)}^{\rm
(part)}$ cancels one of the terms entering $\partial_t {\epsx^i}_{(5)}/c$. 
In the latter case, $(g'_{0i})_{(6)}$
reduces to $-\partial_i {\epsx^0}_{(6)}^{\rm (arb)} + \partial_t
{\epsx^i}_{(5)}^{\rm (arb)}/c$, where ${\epsx^i}_{(5)}^{\rm (arb)}$ 
is an arbitrary function of elementary potentials. We have then
\[ n_d((g'_{0i})_{(6)}) = \max[n_d({\epsx^0}_{(6)}^{\rm
(arb)}), 1 + n_d({\epsx^i}_{(5)}^{\rm (arb)})] \, . \]
Since $n_d((g'_{0i})_{(6)})$ as well as $n_d((g'_{00})_{(7)})$ 
are adjustable, we shall succeed in further lowering the order of temporal
derivatives if $n_d((g'_{ij})_{(5)}) \le 3$. To know
whether this inequality is fulfilled, we solve the equation
$- \partial_i {\epsx^0}_{(6)}^{\rm (part)} + \partial_t {\epsx^i}_{(5)}^{\rm 
(part)}/c=0 $ for ${\epsx^i}_{(5)}^{\rm (part)} \equiv {\epsx^i}_{(5)} -
{\epsx^i}_{(5)}^{\rm (arb)}$. The solution that vanishes when $I_{ij}(t)=0$
reads 
\begin{equation} \label{eq:epsi5_part}
{\epsx^i}_{(5)}^{\rm (part)} = c \, \partial_i {\epsx^{0}}_{(6)}^{{\rm
(part)}(-1)} = \partial_i U_{(5)}^{{\rm reac}(-2)} \, ,
\end{equation}
which leads to odd metric space components of the form $(g'_{ij})_{(5)}
=\partial_{(i} {\epsx^{j)}}_{(5)}^{\rm (arb)} - 2 G I^{(3)}_{ij}(t)$ 
$/ (5c^5)$,
with $n_d((g'_{ij})_{(5)}) = 3$. No function of
${\epsx^0}_{(6)}^{\rm (arb)}$ or ${\epsx^i}_{(5)}^{\rm (arb)}$ having a maximal
time derivative order smaller than $3$ are capable to 
rule out the second term, so that we may take ${\epsx^0}_{(6)}^{\rm (arb)} =
{\epsx^i}_{(5)}^{\rm (arb)} = 0$.

Having specified the gauge vector at the $2.5$ PN order, we can
proceed to the next approximation, and determine both ${\epsx^{0}}_{(8)}$
and ${\epsx^i}_{(7)}$ by using the same method as employed above. The matrix 
$h_{\mu\nu}$, depending on none of the unknown functions, is entirely fixed at
this stage. The highest order temporal derivations again arise in the
reaction potential. At this level, we have 
$n_d((g'_{\mu\nu})_{\rm odd}) = n_d(U^{\rm reac}_{(7)}) =
7$. As before, we intend to make this number decrease by discarding the term
$U^{\rm reac}_{(7)}$ from the metric \eqref{eq:gnew}. Its
complete or partial absorption into ${\epsx^i}_{(7)}$ would result in a
explicit dependence:
${\epsx^i}_{(7)}={\epsx^i}_{(7)}(U^{\rm reac}_{(7)})$, which
would entail the appearance of time derivatives of seventh
order in the spatial part \eqref{eq:gijnew} of $g'_{\mu\nu}$. Considering that
$(g'_{ij})_{(7)}-2 \partial_{(i} {\epsx^{j)}}_{(7)}$ contains fifth order
derivatives at most, these high order time derivatives could not be erased by
means of other contributions. We would end up to the same value of
$n_d((g'_{\mu\nu})_{\rm odd})$ as before:
$n_d((g'_{\mu\nu})_{\rm odd}) = n_d((g'_{ij})_{(7)}) =
7$. Because it is not acceptable, we must rather incorporate 
$U^{\rm reac}_{(7)}$ into ${\epsx^0}_{(8)}$, and set
\begin{multline} \label{eq:eps08_part_arb}
{\epsx^0}_{(8)} = \frac{1}{c} U^{{\rm reac}(-1)}_{(7)} +
{\epsx^0}_{(8)}^{\rm (arb)}  \, ,\\
n_d({\epsx^0}_{(8)}^{\rm (arb)}) \le 6 \, ,
\end{multline}
$n_d((g'_{\mu\nu})_{\rm odd})$ going automatically down 
to $6$. The highest order time derivatives actually
come from the terms $-\partial_i {\epsx^0}_{(8)}$ and $-4 U_{i(5)}^{\rm
reac}/c^3$ in $(g'_{0i})_{(8)}$, but they can be eliminated if
a part of $-\partial_i {\epsx^0}_{(8)}^{\rm (arb)} + \partial_t
{\epsx^i}_{(7)}$ exactly cancel them. This happens for
\begin{multline} \label{eq:epsi7_part_arb}
{\epsx^i}_{(7)} = \partial_i U_{(7)}^{{\rm reac}(-2)} + \frac{4}{c^2}
U_{i(5)}^{{\rm reac}(-1)} + {\epsx^i}_{(7)}^{\rm (arb)} \, , \\
n_d({\epsx^i}_{(7)}^{\rm (arb)}) \le 5 \, .
\end{multline}
The space components of the new $3.5$ PN metric are obtained by inserting
the expression above into the right-hand side of equation \eqref{eq:gijnew}.
This yields a maximal order of temporal derivations equal to
$n_d({\epsx^i}_{(7)}) =5$, whatever functions
${\epsx^0}_{(8)}^{\rm (arb)}$ and ${\epsx^i}_{(7)}^{\rm (arb)}$ we have used.
We have still the freedom to specify the ``arbitrary'' part of $\epsx$ as
most convenient for us.

In summary, it is possible to reduce the number $n_d(g'_{\mu\nu})$ from $7$ to
$5$ by performing a suitable (linear) gauge transformation. Equations
\eqref{eq:eps06_part_arb}, \eqref{eq:epsi5_part},
\eqref{eq:eps08_part_arb}, and \eqref{eq:epsi7_part_arb}
provide a possible choice for the gauge vector:
\begin{subequations}\label{eq:eps}
\begin{align}
{\epsx^0}_{(6+8)} =&~ \frac{1}{c} \partial_t U^{{\rm reac}(-2)}_{(5+7)} \, ,
\nonumber \\ {\epsx^i}_{(5+7)} =&~ \partial_i U^{{\rm reac}(-2)}_{(5+7)}
+ \frac{4}{c^2} U^{{\rm reac}(-1)}_{i(5)} \, ,
\end{align}
with
\begin{align*}
U^{{\rm reac}(-2)} &= - \frac{G}{5c^5} x^{ij} I^{(3)}_{ij}(t)
+ \frac{G}{c^7} \bigg[ \frac{1}{189} x^{ijk} I^{(5)}_{ijk}(t)
  \nonumber \\ &  -
\frac{1}{70} x^{kk} x^{ij} I^{(5)}_{ij}(t) \bigg] + \cO
\left(\frac{1}{c^8}\right)
\end{align*}
and
\begin{align}
U^{{\rm reac}(-1)}_i =&~ \frac{G}{c^5} \bigg[\frac{1}{21}
\hat{x}^{ijk} I^{(5)}_{jk}(t) \nonumber \\ & - \frac{4}{45} \epsilon_{ijk}
x^{jm} J^{(4)}_{km}(t) \bigg] + \cO\left( \frac{1}{c^6} \right)
\end{align}
\end{subequations}
modulo the ``arbitrary'' part.
There exist other interesting alternatives. For instance, we could let the
original metric unchanged at the $2.5$ PN level while keeping the
requirement $n_d((g'_{\mu\nu})_{\rm odd}) \le 5$ since neither 
$(g_{00})_{(7)}$, $(g_{0i})_{(6)}$, or $(g_{ij})_{(5)}$ involve time
derivatives of higher order. The coordinate system defined by
means of formulas \eqref{eq:eps} present the advantage to coincide with the  
ADM one up to the $2.5$ PN approximation, as we shall see in the next section.

\section{Expression of the odd metric at the $\mathbf{3.5}$ PN order}

For computing the reaction force, we must now finalize the form of
the gravitational field by fixing the two arbitrary functions
${\epsx^0}_{(8)}^{\rm (arb)}$ and ${\epsx^i}_{(7)}^{\rm (arb)}$. On one hand,
the gauge $x'^{\mu} = x^{\mu} - \epsx^\mu$ is constructed in such a way
that $g'_{\mu\nu}$ is precisely the ADM metric $(g_{\mu\nu})_{\rm ADM}$ at the
1 PN order: $g'_{\mu\nu} = (g_{\mu\nu})_{\rm ADM}$, plus a 2 PN deviation. On
the other hand, $(g'_{00})_{(9)}$ depends on the 1 PN components of the metric
\eqref{eq:metric} in the Burke-Thorne-like gauge introduced in paper
\cite{B97}. 
Now, the non-spatial terms of $(g_{\mu\nu})_{\rm ADM}$ differ from
those of $g_{\mu\nu}$ at this order. We have:
\begin{align}
(g_{00})_{{\rm ADM}\, (4)} =& - \frac{2}{c^4} U^2 \, , \nonumber \\ 
(g_{0i})_{{\rm ADM}\, (3)} =& -\frac{1}{c^3} A_i \equiv -\frac{1}{c^3} \left(4
U_i + \frac{1}{2} \partial_i \partial_t \chi \right)  \, .
\end{align}
In order to homogenize the expression of the $3.5$ PN metric, we shall choose
the ``arbitrary'' part of $\epsx$ so as to discard the contribution of
$(g_{00})_{(4)}$ (resp. $(g_{0i})_{(3)}$) and replace it by that of
$(g_{00})_{{\rm ADM}\, (4)}$ (resp. $(g_{0i})_{{\rm ADM} \, (3)})$ in
$(\delta^* g'_{\mu\nu})_{\rm odd}$. This is achieved by taking
\begin{align*}
{\epsx^0}_{(8)}^{\rm (arb)} =&~ {\epsx^i}_{(5)} \partial_i {\epsx^0}_{(3)}
- D_{X_B}[{\epsx^0}_{(3)}] \cdot \left(\delta^*_{\varepsilon^\beta_{\rm
odd}} \! \! \! \! X_B \right)_{(5)} \\ =&~
{\epsx^0}_{(6)} \partial_0 {\epsx^0}_{(3)} + {\epsx^i}_{(5)} \partial_i
{\epsx^0}_{(3)} \\ & - D_{X_B}[{\epsx^0}_{(3)}] \cdot
\delta^*_{\varepsilon^\beta_{\rm odd}} \! \! \! \! X_B + \cO \left(
\frac{1}{c^{10}} \right) \, , \\ 
{\epsx^i}_{(7)}^{\rm (arb)} =&\; 0 = {\epsx^0}_{(6)} \partial_0
{\epsx^i}_{(4)} + {\epsx^j}_{(5)} \partial_j {\epsx^i}_{(4)} \nonumber \\
&\quad - D_{X_B}[{\epsx^i}_{(4)}] \cdot \delta^*_{\varepsilon^\beta_{\rm odd}}
\! \! \! \! X_B + \cO \left( \frac{1}{c^9} \right) \, ,
\end{align*}
with $\epsilon^\mu_{\rm odd} \equiv ({\epsx^0}_{(6+8)},{\epsx^i}_{(5+7)})$.
This statement can be checked either directly or by use of formula
\eqref{eq:x_xpp} of section \ref{sec:transformation}.
noticing that the final value of $\epsx$, which reads
\begin{equation} \label{eq:epsilon_square}
 \epsx^\mu = \varepsilon^\mu_{\rm even} + \varepsilon^\mu_{\rm odd} +
\varepsilon^\lambda_{\rm odd} \partial_\lambda \varepsilon^\mu_{\rm even}-
D_{X_B}[\varepsilon^\mu_{\rm even}] \cdot \delta^*_{\varepsilon^\beta_{\rm
odd}}\! \! X_B \end{equation}
if we pose $\varepsilon^\mu_{\rm even} = ({\epsx^0}_{(3)}, {\epsx^i}_{(4)})$,
is exactly the one we would get by applying successively the two coordinate
transformations $x''^\mu = x^\mu - \varepsilon^\mu_{\rm even}(x''^\alpha,
X''_A)$ and $x'^\mu = x''^\mu - \varepsilon^\mu_{\rm odd} (x'^\alpha,X'_A)$.

The resulting metric $g'_{\mu\nu}$ is
\begin{subequations} \label{eq:gmn}
\begin{align}
g'_{00} =& -1 + \frac{2}{c^2} U - \frac{2}{c^4} U^2 - \frac{4}{c^4} U U^{\rm
reac} + \frac{4}{c^3} U 
\partial_t {\epsx^0}_{(6)} \nonumber \\ & - \frac{2}{c^4}
A_i \partial_t {\epsx^i}_{(5)} + \frac{2}{c^3} {\epsx^0}_{(6)}
\partial_t U + \frac{1}{c^4} {\epsx^i}_{(5)} \partial_i (-2
U^2)  \nonumber \\ & +  \frac{2}{c^2} {\epsx^i}_{(5+7)} \partial_i U -
\frac{2}{c^2} \integ  
d^3\mathbf{y}~\frac{\delta U}{\delta X_A} (\delta^*_{\epsx}
X_A)_{(5+7)}  \nonumber \\ & - \frac{1}{c^4} \integ d^3\mathbf{y}~\frac{\delta
(-2 U^2)}{\delta X_A} (\delta^*_{\epsx} X_A)_{(5)} +
\frac{1}{c^6} (\ldots) \nonumber \\ & + 
\frac{1}{c^8} (\ldots) + \cO \left( \frac{1}{c^{10}} \right)\, , \\
g'_{0i} =& -\frac{1}{c^3} A_i  
+ \frac{2}{c^2} U \partial_i {\epsx^0}_{(6)} + \frac{2}{c^3} U \partial_t
{\epsx^i}_{(5)} \nonumber \\ & - \frac{1}{c^3} A_j \partial_i
{\epsx^j}_{(5)} - \frac{1}{c^3} {\epsx^j}_{(5)} \partial_j A_i 
\nonumber \\ & + \frac{1}{c^3} \integ 
d^3\mathbf{y}~\frac{\delta A_i}{\delta X_A} (\delta^*_{\epsx}
X_A)_{(5)} + \frac{1}{c^5} (\ldots) \nonumber \\ & + \frac{1}{c^7} (\ldots)
+ \cO \left( \frac{1}{c^9} \right)\, , \\
g'_{ij} =&~ \delta^{ij} \left[1 + \frac{2}{c^2} U + \frac{2}{c^2} U^{\rm
reac}_{(5)} \right] + 2 \partial_{(i} {\epsx^{j)}}_{(5+7)} \nonumber \\ & +
\frac{4}{c^2} U \partial_{(i} 
{\epsx^{j)}}_{(5)} + \frac{2}{c^2} {\epsx^k}_{(5)}  \partial_k U \delta^{ij}
\nonumber \\ & - \frac{2}{c^2} \delta^{ij} \integ d^3\mathbf{y}~\frac{\delta
U}{\delta X_A}  (\delta^*_{\epsx} X_A)_{(5)} + \frac{1}{c^4} (\ldots)
\nonumber \\ & + \frac{1}{c^6} (\ldots) + \cO \left( \frac{1}{c^8} \right)\, ,
\end{align}
\end{subequations}
where ${\epsx^0}_{(6+8)}$ and ${\epsx^i}_{(5+7)}$ are defined in
equations \eqref{eq:eps}. We remind that $(g'_{0i})_{(6)}=0$ 
by construction. We may now calculate
the functional derivatives explicitly. Considering for
example the potential $U$, its source contains none of the
densities $\sigma_i$ or $\sigma_{ij}$, and thus, $\integ d^3\mathbf{y}\,
(\delta U/\delta \sigma_i)\,\delta^* \sigma_i = \integ d^3\mathbf{y} \,
(\delta U/\delta \sigma_{ij})\, \delta^* \sigma_{ij} = 0$. The term $\integ
d^3\mathbf{y} \, (\delta U/\delta \sigma) \, \delta^* \sigma$ is deduced from
the evaluation of the difference $U'-U$:
\begin{align*}
U'- U &= G \integ \frac{d^3\mathbf{y}}{|\mathbf{x} - \mathbf{y}|}~
\sigma'(\mathbf{y}, t) - G \integ \frac{d^3\mathbf{y}}{|\mathbf{x} -
  \mathbf{y}|}~ \sigma(\mathbf{y}, t) \\ &=  G \integ
\frac{d^3\mathbf{y}}{|\mathbf{x} - \mathbf{y}|}~ \delta^* \sigma(\mathbf{y},
t) + \cO([\delta^* \sigma]^2) \, ,
\end{align*}
hence
\begin{subequations} \label{eq:frechetP}
\begin{align} \label{eq:frechetU_sigma}
\integ &d^3\mathbf{y}~ \frac{\delta U}{\delta \sigma} (\delta^*_{\epsx} 
\sigma)_{(5+7)} \\ & 
= G \integ \frac{d^3\mathbf{y}}{|\mathbf{x} - \mathbf{y}|}~
(\delta^*_{\epsx} \sigma)_{(5+7)} \nonumber \\
&= G \integ \frac{d^3\mathbf{y}}{|\mathbf{x} - \mathbf{y}|}~  \bigg[
{\epsx^k}_{(5+7)} \partial_k \sigma - \frac{2}{c} \sigma \partial_t
{\epsx^0}_{(6)} + \frac{1}{c} {\epsx^0}_{(6)} \partial_t \sigma \nonumber \\ &
- \frac{2}{c} \sigma_i \partial_i {\epsx^0}_{(6)} - \frac{2}{c^2} \sigma_i
\partial_t {\epsx^i}_{(5)} - \frac{2}{c^2} \sigma_{ik} \partial_k
{\epsx^i}_{(5)}  \bigg](\mathbf{y}, t) \, .
\end{align}
The other non-zero functional derivatives present in the metric variation
\eqref{eq:deltag} are computed following the same procedure.
\begin{align}
\integ &d^3\mathbf{y}~ \frac{\delta A_i}{\delta \sigma}
(\delta^*_{\epsx}\sigma)_{(5)} \nonumber \\ &= 4 \integ d^3\mathbf{y}~
\frac{\delta  U_i}{\delta 
\sigma} (\delta^*_{\epsx}\sigma)_{(5)} + \frac{1}{2} \integ d^3\mathbf{y}~
\frac{\delta (\partial_t \partial_i \chi)}{\delta \sigma}
(\delta^*_{\epsx}\sigma)_{(5)} \nonumber \\ &=
\frac{G}{2} \partial_t \partial_i \integ d^3\mathbf{y}~ |\mathbf{x} -
\mathbf{y}| ~[{\epsx^k}_{(5)} \partial_k \sigma](\mathbf{y}, t)\, ,
\label{eq:frechetAi_sigma} \\  
\integ &d^3\mathbf{y}~ \frac{\delta A_i}{\delta \sigma_j}
(\delta^*_{\epsx}\sigma_j)_{(5)} \nonumber \\ & = 4 G
\integ \frac{d^3\mathbf{y}}{|\mathbf{x} - \mathbf{y}|}~ \bigg[- \sigma
\partial_t {\epsx^i}_{(5)} - \sigma_j \partial_j {\epsx^i}_{(5)} \nonumber \\
& \quad \qquad \qquad \qquad \; \; + {\epsx^j}_{(5)} \partial_j  \sigma_i  
\bigg](\mathbf{y}, t) \, . \label{eq:frechetAi_sigmaj}
\end{align}
\end{subequations}

By inserting formulas \eqref{eq:frechetP} into equation \eqref{eq:gmn}, we
get an expression for the metric already well adapted to our
purpose. Notably, all potentials elementary potentials but $\chi$ and $G
\integ d^3 \mathbf{y}~|\mathbf{x}-\mathbf{y}| (\delta \chi/\delta \sigma)
\delta^*_{\epsx} \sigma$ satisfy Poisson-type equations. Nonetheless, in order
to make easier the 
comparison with paper \cite{RSABS99}, we shall present an alternative
form where the functional
derivatives are re-written as combinations of Poisson-type integrals on the
sources $X_1=\sigma$ 
and $X_2=\sigma_i$ exclusively. It can be obtained from an
appropriate handling of the integrands in the potentials \eqref{eq:frechetP}. 
Originally, they are indeed made of sums of elementary pieces of the kind 
\begin{multline*}
|\mathbf{x} - \mathbf{y}|^{2p-1} \partial X_A(\mathbf{y},t) \partial
\varepsilon_{\rm odd}\\
\propto |\mathbf{x} - \mathbf{y}|^{2p-1}  y^{i_1 i_2 \ldots i_l} \partial
X_A({\bf y},t) \times \textrm{function of time},
\end{multline*}
for $A=1,2$ and $p$, $l \in \mathbb{N}$. Now,
it is straightforward to show by recurrence that 
\begin{align*}
&y^{i_1 i_2 \ldots i_l} =  x^{i_1 i_2 \ldots i_l} - [x^{i_1 i_2 \ldots
i_{l-1}} (x^{i_l}-y^{i_l}) + \textrm{permutations}]\\ & \quad +
 [x^{i_1 i_2 \ldots i_{l-2}} (x^{i_{l-1}}-y^{i_{l-1}}) (x^{i_l}-y^{i_l}) +
\textrm{permutations}] \\ & \quad + \ldots 
+ (-1)^l (x^{i_1}-y^{i_1}) (x^{i_2}-y^{i_2}) \ldots (x^{i_l}-y^{i_l}) \, ,
\end{align*}
so that every piece decomposes into a sum of terms such as
$x^{j_1 j_2 \ldots j_k} (x^{i_1} - y^{i_1}) (x^{i_2} - y^{i_2}) \ldots
(x^{i_l} - y^{i_l})|\mathbf{x} - \mathbf{y}|^{2p-1}$. Each of them are next
transformed by means of the identity
\begin{widetext}
\[
(x^{i_1} - y^{i_1}) \ldots
(x^{i_l} - y^{i_l})|\mathbf{x} - \mathbf{y}|^a = 
\sum^{[l/2]}_{m=0} \left[ \frac{(-1)^m  \delta^{i_1 i_2} \ldots
    \delta^{i_{2m-1} i_{2m}}}{(a+2l-2m) \ldots (a+2)} 
\partial_{i_{2m+1}
\ldots i_l} |\mathbf{x} - \mathbf{y}|^{a+2l-2m}+ \textrm{permutations}
\right]\, , \]
valid for $a > -2$, where the latter relation results from the formula
\begin{multline*}
\partial_{i_1 i_2 \ldots i_l} |\mathbf{x} - \mathbf{y}|^a = \sum_{m=0}^{[l/2]}
a (a-2) \ldots (a-2l+2m+2) |\mathbf{x} -\mathbf{y}|^{a-2l+2m}
[\delta^{i_1 i_2} \ldots \delta^{i_{2m-1} i_{2m}} \\ \times (x^{i_{2m+1}} -
y^{i_{2m+1}}) \ldots (x^{i_l} - y^{i_l}) + \textrm{permutations}] \, ,
\end{multline*}
\end{widetext}
proved by simple recurrence. The square bracket $[l/2]$ denotes here the
integer part of $l/2$. We thus see that the initial potentials can be
split into a sum of terms matching the pattern: 
\begin{multline*}
\integ d^3\mathbf{y}~ x^{j_1 j_2 \ldots j_k} \partial_{i_1 i_2 \ldots i_l}
|\mathbf{x} - \mathbf{y}|^{2p-1} \partial X_A(\mathbf{y},t) \\= x^{j_1 j_2
  \ldots j_k} \partial \partial_{i_1 i_2 \ldots i_l} \integ d^3\mathbf{y}~ 
|\mathbf{x} - \mathbf{y}|^{2p-1} X_A(\mathbf{y},t) \, .
\end{multline*}
Under this form, they only depend on the Poisson-like integrals $P_{q} =
G \integ d^3\mathbf{y}~|\mathbf{x} - \mathbf{y}|^{2q-1} \sigma(\mathbf{y},
t)$, $P_{iq} = G \integ d^3\mathbf{y}~|\mathbf{x} - \mathbf{y}|^{2q-1}
\sigma_i(\mathbf{y}, t)$ as well as 
$P_{ijq} = G \integ d^3\mathbf{y}~|\mathbf{x} - \mathbf{y}|^{2q-1}
\sigma_{ij}(\mathbf{y}, t)$. We shall need in particular $P_{0} = U$,
$P_{1}=\chi$, $P_{i0}=U_i$ as before, but also $\chi_i = P_{i1}$ and $U_{ij} =
P_{ij0}$. As an example, let us apply the latter treatment to the simplest
functional derivative $G \integ d^3\mathbf{y}/|\mathbf{x} - \mathbf{y}|~
{\epsx^k}_{(5)} \partial_k \sigma$. We have: 
\begin{align*}
G \integ & \frac{d^3\mathbf{y}}{|\mathbf{x} - \mathbf{y}|}~ 
{\epsx^k}_{(5)} \partial_k \sigma \\ & = 
G \integ \frac{d^3\mathbf{y}}{|\mathbf{x} - \mathbf{y}|} \left(-\frac{2G}{5
    c^5}  y^{i} I^{(3)}_{ij}(t)\right) \partial_j \sigma \\ &=
- \frac{2 G^2}{5 c^5} I^{(3)}_{ij}(t) \integ d^3\mathbf{y}
\left[-\frac{(x^i - y^i)}{|\mathbf{x} - \mathbf{y}|} + \frac{x^i}{|\mathbf{x}
    - {\bf y}|}  \right] \partial_j \sigma \\
& =- \frac{2 G^2}{5c^5} I^{(3)}_{ij}(t)\bigg[- \integ d^3\mathbf{y}
\frac{\partial}{\partial x^i} |\mathbf{x} - \mathbf{y}|
\frac{\partial}{\partial y_j} \sigma \\ & \qquad \qquad \qquad \quad + x^i
\integ \frac{d^3{\bf y}}{|\mathbf{x} - \mathbf{y}|} \frac{\partial
  \sigma}{\partial y_j}\bigg] \, . 
\end{align*}
After integrating by part and observing that $\partial |\mathbf{x} - {\bf
y}|^a/\partial y^i =-\partial |\mathbf{x} - \mathbf{y}|^a/\partial x^i$ for
any real number $a$, we arrive at the equality: $G \integ d^3\mathbf{y}~
{\epsx^k}_{(5)} \partial_k \sigma/|\mathbf{x} - \mathbf{y}| =- 
2G I^{(3)}_{ij}(t) [-\partial_{ij} \chi$ $ + x^i \partial_j  U ]/(5c^5)$. 
Processing all the elementary potentials in a similar way leads to the
important result
\begin{align*}
(&g'_{00})_{\rm odd} = - 
     \frac{4 G}{5 c^7} I_{kl}^{(3)} \partial_{kl} \chi\\ & + \frac{2G}{c^9}
     \bigg( 
     \frac{4}{5} I_{kl}^{(3)} (U \partial_{kl} \chi - U_{kl}) \\ & \qquad +
     \frac{1}{5} I_{kl}^{(4)} (8 \partial_k \chi_l + 
     \frac{1}{3} \partial_{kl} \partial_t P_2 - 
     x^k \partial_l \partial_t \chi)\\ & \qquad + 
     \frac{1}{7} I_{kl}^{(5)} ( - 
     \frac{8}{15} x^k x^l U -
     \frac{11}{15}  r^2 \partial_{kl} \chi - 
     \frac{29}{15} x^k \partial_l \chi \\ & \qquad \qquad \qquad \! + 
     \frac{4}{5} x^k x^m \partial_{ml} \chi +
     \frac{1}{3} \partial_{kl} P_2 + 
     \frac{1}{9} x^m \partial_{klm} P_2 )\\ & \qquad + 
     \frac{1}{63} I_{klm}^{(5)} (-\frac{1}{3} \partial_{klm} P_2 +
     2 x^k \partial_{lm} \chi) \\ & \qquad - 
     \frac{16}{45} x^k \epsilon_{klm} J_{ln}^{(4)} \partial_{mn} \chi \bigg)
     \, ,\\
(&g'_{0i})_{\rm odd} = \frac{G}{5c^8} \bigg(
     16 I_{ik}^{(3)} U_k + I_{kl}^{(3)} (  
     8 \partial_{kl} \chi_i + 
     \frac{1}{3} \partial_{ikl} \partial_t P_2) \\ & \qquad \qquad \qquad - 
     9 I_{ik}^{(4)} \partial_k \chi + I_{kl}^{(4)} ( - 
     x^k \partial_{il} \chi +
     \frac{1}{3} \partial_{ikl} P_2 ) \bigg) \, ,\\
(&g'_{ij})_{\rm odd} = -\frac{4 G}{5 c^5} I_{ij}^{(3)} \\ & + 
     \frac{2G}{c^7} \bigg(- 
     \frac{4}{5} I_{ij}^{(3)} U -
     \frac{2}{5} \delta^{ij} I_{kl}^{(3)}  \partial_{kl} \chi - 
     \frac{11}{105} r^2 I_{ij}^{(5)} \\ & \qquad \qquad + 
     \frac{2}{63} x^k I_{ijk}^{(5)} + 
     \frac{4}{35} x^k x^{(i} I_{j)k}^{(5)} \\ & \qquad \qquad - 
     \frac{4}{105} \delta^{ij} x^k x^l I_{kl}^{(5)} - 
     \frac{16}{45} x^k \epsilon_{kl(i} J_{j)l}^{(4)} \bigg) \, ,
\end{align*}
with $r^2 = x^m x^m$. The leading components of $(g'_{\mu\nu})_{\rm odd}$ are
identical to those of the $2.5$ PN metric in ADM coordinates. With the help of
these formulas, we recover the expression proposed by Rezzolla
\emph{et al.} in paper 
\cite{RSABS99} in absence of mass multipole moments, i.e. when $I_{i_1
i_2 \ldots i_l}=0$. We conclude that our gauge coincides with the ADM one up
to the $2.5$ PN order, and generalizes that of Rezzolla and collaborators
at the next level. 

Once the gravitational field has been evaluated in the new coordinate system,
we may take advantage of the freedom in the choice of the matter
variables. Variables defined by means of the fluid trajectory in the
configuration space without any reference to the metric are particularly
appropriate to post-Newtonian calculations, since their Poisson brackets are
identical to those of Newtonian theory \cite{H85}. We first take
the baryonic density $\rhoc$ representing the number of
baryon weighed by their individual mass per \emph{coordinate} volume $d^3{\bf
x}$. It reduces to the scalar density $\rho$ at the Newtonian order: $\rhoc =
\sqrt{-g'} \rho u^0 = \rho + \cO(1/c^2)$. The second variable we shall use
for describing 
the fluid is the coordinate velocity $v^i$. It does not depend on
the metric as being the Eulerian quantity associated to the
Lagrangian velocity $dx^\mu(x^\alpha_0,t)/dt$ of the particle
$x^\mu(x^\alpha_0,t)$ located at point $x^i_0$ at time $x^0_0/c$. We have
$v^i/c = u^i/u^0$, and therefore, $v^i/c = u^i + \cO(1/c^2)$.
We shall introduce a zeroth component $v^0 = c$ when necessary, so that we can
write $v^\mu/c = u^\mu/u^0$. For macroscopic systems,
the set of variables $\{\rhoc, v^i\}$ is completed by the entropy $s$.

The relation between $\sigma$ and
$\sigma_i$ on one hand, $\rhoc$, $v^i$, and $s$ on the other hand, is
determined by the form of the stress-energy tensor. We shall focus
henceforth on \emph{adiabatic fluid} systems, for which
\begin{equation} \label{eq:Tmn}
T^{\mu\nu} = \rho (c^2 + h) u^{\mu\nu} + p g'^{\mu\nu} \, ,
\end{equation}
where $h$ and $p$ stand for the enthalpy per unit mass and for the pressure
respectively. The stress-energy tensor \eqref{eq:Tmn} has to
be supplemented by the equation of state of the fluid; it is typically
provided by the energy $e$ per 
unit mass\footnote{Let us point out that the symbol $e$ denotes the energy
per volume unit in paper \cite{BDS90}} as a
function of the scalar density and of the entropy, $e = e(\rho,s)$, from what
we can infer the pressure $p(\rho,s)=\rho^2 \partial e(\rho,s)/\partial
\rho$ or the enthalpy $h(\rho,s) = e(\rho,s) + p(\rho,s)/\rho$.
It is now possible to relate the densities $\sigma$ to $\rhoc$, $v^i$, and $s$.
At the Newtonian approximation, we have $\sigma =
(T^{00}+T^{ii})/c^2 = [\rho c^2 {u^0}^2 + \cO (1/c^0)]/c^2 = \rhoc +
\cO(1/c^2)$, and $\sigma_i= T^{0i}/c = [\rho c^2 u^i u^0 + \cO(1/c)]/c = \rhoc
v^i + \cO(1/c^2)$. It is thus useful to pose:
\begin{subequations}
\begin{align}
\Uc &= G \integ \frac{d^3\mathbf{y}}{|\mathbf{x} - \mathbf{y}|}~ \rhoc \, , \\
\Uc_i &= G \integ \frac{d^3\mathbf{y}}{|\mathbf{x} - \mathbf{y}|}~ \rhoc v^i ,
\\ \chi_* &= G \integ d^3\mathbf{y} |\mathbf{x} - \mathbf{y}| ~ \rhoc \, .
\end{align}
\end{subequations}

At the order 1 PN and beyond we proceed in several steps, for
the computations are longer. We first expand $\sqrt{-g'}$ in
power of $1/c$ in order to evaluate $\rho(\rhoc,v^p,s)=\rhoc/(u^0\sqrt{-g'})$.
From the expression of the determinant of a perturbed Minkowski metric
$g'_{\mu\nu} = \eta_{\mu\nu} + \delta g'_{\mu\nu}$ at linear order, $g'=-1+
\delta g' + \cO((\delta g')^2)$ with $\delta g' = \det(\eta_{\alpha \beta})
\eta^{\rho\sigma} \delta  g'_{\rho \sigma} = -\eta^{\alpha \beta} \delta
g'_{\alpha \beta}$, we get immediately
\begin{align*}
\sqrt{-g'} =&~ (1 + \eta^{\alpha\beta} \delta g'_{\alpha\beta})^{1/2} + 
\cO((\delta g')^2)\\
=&~ 1 - \frac{1}{2} (g'_{00})_{(\le 7)} + \frac{1}{2}
(g'_{ii})_{(\le 7)} + \cO \left(\frac{1}{c^2}\right) 
\cO \left( g'_{\rm odd}\right) \\& + \frac{1}{c^4} (\ldots) + \frac{1}{c^6}
(\dots)+ \cO\left(\frac{1}{c^8}\right) \, .
\end{align*}
The lowest contribution to $\cO(1/c^2) \cO(g'_{\rm odd})$ is a
Lorentz scalar proportional to $(g'_{00})_{(2)}$ or $(g'_{ij})_{(2)}$, and
$(g'_{ij})_{(5)}$. Now, the only allowed combinations are 
$(g'^{\alpha \beta})_{(2)} \times (g'_{\alpha\beta})_{(5)} \propto \delta^{ij} 
{g'_{ij}}_{(5)}$, and $\eta^{\alpha\beta} (g'_{\alpha\beta})_{(2)} \eta^{\rho
\sigma} (g'_{\rho\sigma})_{(5)}$, which are both zero due to the
trace-free property of $(g'_{ij})_{(5)} = -4 G I^{(3)}_{ij}/(5c^5)$. As a
consequence,
\begin{multline} \label{eq:volume}
\sqrt{-g'} = 1 +
\frac{2}{c^2} U - \frac{1}{2} (g'_{00})_{(7)} + \frac{1}{2} (g'_{ii})_{(7)} +  
\frac{1}{c^4}( \ldots ) \\ + \frac{1}{c^6} ( \ldots ) + \cO
\left(\frac{1}{c^8}
\right) \, .
\end{multline}
We next calculate the Lorentz factor $u^0 = dt/d\tau$ appearing in
$\rho$ and $v^i$, from the definition of the proper time $c d\tau =
(-g'_{\alpha \beta} dx^\alpha dx^\beta)^{1/2}$. We find:
\begin{align} \label{eq:u0_intermediate}
u^0 =&~ \left[- g'_{00} - 2 g'_{0i} \frac{v^i}{c} -
g'_{ij} \frac{v^i v^j}{c^2} \right]^{-1/2} \nonumber \\ 
=&~ \left[1 - \frac{2}{c^2} U -
(g'_{00})_{(7)} - 2 (g'_{0i})_{(6)} \frac{v^i}{c} - \frac{v^2}{c^2} \right.
\nonumber \\  & \left. \,  - (g'_{ij})_{(5)} \frac{v^i v^j}{c^2}+
  \frac{1}{c^4} ( \ldots ) + \frac{1}{c^6} ( \ldots ) + \cO 
\left(\frac{1}{c^8} \right)\right]^{-1/2} \nonumber \\ 
=&~ 1 + \frac{U}{c^2} + \frac{1}{2} \frac{v^2}{c^2} +
\frac{1}{2} (g'_{00})_{(7)} + \frac{1}{2}
(g'_{ij})_{(5)} \frac{v^i v^j}{c^2} \nonumber \\ 
& + \frac{1}{c^4} ( \ldots ) + \frac{1}{c^6} ( \ldots ) + 
\cO \left(\frac{1}{c^8} \right) \, ,
\end{align}
where $v$ denotes the Euclidean norm of $v^i$, i.e. $v^2 = v^i v^i$.
Regarding the fact that the lowest odd order part of the matter or
thermodynamical quantities involved in the stress-energy tensor \eqref{eq:Tmn}
is necessarily larger than five, equations \eqref{eq:volume} and
\eqref{eq:u0_intermediate} show that the first odd terms in power
of $1/c$ entering the $1/c$ expansion of $u^0= u^0(\rhoc, v^p, s)$ and
$\rho=\rho(\rhoc, v^p, s)$ arise exactly at order seven, which implies
that the $1/c^5$ parts of the 
scalar density and the Lorentz factor are both zero. Referring to these two
parts 
as $u^0_{(5) \, {\rm reac}}$ and $\rho_{(5) \, {\rm reac}}$ respectively, so as
to distinguish them from the fifth order of $\rho$ and $u^i$ considered as
functions of $\sigma$, $\sigma_i$, and $\sigma_{ij}$, we can write in short:
$u^0_{(5) \, {\rm reac}} = \rho_{(5) \, {\rm reac}} =0$. Therefore, the
densities $\sigma = \rho (u^0)^2+ \cO(1/c^2)$ and $\sigma_i = \rho (u^0)^2
v^i+ \cO(1/c^2)$ 
have no contribution of order $1/c^5$, i.e.
$\sigma_{(5) \, {\rm reac}}=0$ and $\sigma_{i(5) \, {\rm reac}}=0$ (with the
same notation as before). Let us mention that
the stress density $\sigma_{ij} = \rho (u^0)^2 v^i v^j + p g'^{ij}
+ \cO(1/c^2)$ does contribute at this level:
\begin{align} \label{eq:sigmaijreac}
\sigma_{(5)ij \, {\rm reac}} &= p(\rhoc, s) (g'^{ij})_{(5) \, {\rm reac}}
\nonumber \\ &= - p(\rhoc,s) (g'_{ij})_{(5) \, {\rm reac}} \neq 0 \, .
\end{align}
The fact that we have $U_{(5) \, {\rm reac}} = \chi_{(5) \, {\rm
reac}} = 0$ and $U_{i(5) \, {\rm reac}} = 0$ for the potentials leads us to
conclude that the reaction part of the gravitational field is simply
\begin{subequations} \label{eq:gmn2.5PN}
\begin{align} 
(g'_{00})_{(7) \, {\rm reac}} &= \bigg[-1 + \frac{2}{c^2} U + 
\frac{1}{c^4} ( \ldots ) \nonumber \\ & \qquad + \frac{1}{c^6} ( \ldots )+ 
(g'_{00})_{(7)} \bigg]_{\genfrac{}{}{0pt}{1}{(7)}{\rm reac}} = 
[(g'_{00})_{(7)}]_{\genfrac{}{}{0pt}{1}{(7)}{\rm reac}}  \nonumber \\ & =
\frac{4 G}{5c^7} \bigg(- Q_{kl}^{(3)} x^k \partial_l  \Uc  \nonumber \\ &
\qquad \qquad +
G \integ \frac{d^3\mathbf{y}}{|\mathbf{x} - \mathbf{y}|}~Q_{kl}^{(3)} x^k
\partial_l \rhoc \bigg) \, ,\\
(g'_{0i})_{(6) \, {\rm reac}} & = \bigg[-\frac{1}{c^3} \left(4 U_i + 
\frac{1}{2} \partial_i \partial_t \chi \right) \nonumber \\ & \qquad +
\frac{1}{c^5} ( \ldots ) +  (g'_{0i})_{(6)} \bigg]_{(6) \, {\rm reac}} 
= 0 \, ,\\ 
(g'_{ij})_{(5) \, {\rm reac}} &= \left[ \delta^{ij} \left(1 + \frac{2}{c^2} U
\right) + \frac{1}{c^4} ( \ldots ) + (g'_{ij})_{(5)}
\right]_{\genfrac{}{}{0pt}{1}{(5)}{\rm reac}} \nonumber \\ & =
[(g'_{ij})_{(5)}]_{(5) \,  {\rm reac}} = - \frac{4 G}{5c^5} Q_{ij}^{(3)}
\end{align}
\end{subequations}
at the leading approximation. The functions $[(g'_{00})_{(7)}]_{(7) \, {\rm
reac}}$ and $[(g'_{ij})_{(5)}]_{(5) \, {\rm reac}}$, are obtained by replacing
consistently the Newtonian-like potentials $U$, $U_i$, $\chi$, and the mass
quadrupole moment $I_{ij}$, by $\Uc$, $\Uc_i$, $\chi_*$, and $Q_{ij} = \integ
d^3\mathbf{y}~ \hat{y}^{ij} \rhoc$ respectively in the corresponding odd metric
components. From equations \eqref{eq:volume},
\eqref{eq:u0_intermediate}, and \eqref{eq:gmn2.5PN},
we deduce $u^0$ and $\rho$ expressed within our set of
variables:
\begin{align}
u^0 &=  1 + \frac{\Uc}{c^2} + \frac{1}{2} \frac{v^2}{c^2} +
\frac{1}{2} [(g'_{00})_{(7)}]_{(7) \, {\rm reac}}
\nonumber \\ 
&  + \frac{1}{2} [(g'_{ij})_{(5)}]_{\genfrac{}{}{0pt}{1}{(5)}{\rm reac}}
\frac{v^i v^j}{c^2} + 
\frac{1}{c^4} ( \ldots ) + \frac{1}{c^6} ( \ldots ) +  
\cO \left(\frac{1}{c^8} \right) \, ,  \label{eq:u0} \\
\rho &= \rhoc \left( 1 - \frac{3}{c^2} \Uc -
\frac{1}{2} \frac{v^2}{c^2} \right) \nonumber \\ & - \frac{\rhoc}{2} \left(
[(g'_{ii})_{(7)}]_{(7) {\rm reac}} + 
[(g'_{ij})_{(5)}]_{(5) \, {\rm reac}} \frac{v^i v^j}{c^2}  \right) \nonumber \\
& + \frac{1}{c^4} ( \ldots ) + \frac{1}{c^6} ( \ldots ) + \cO 
\left( \frac{1}{c^8} \right) \, . \label{eq:rho}
\end{align}
At last, we compute the pressure $p$ as a function of $\rhoc$, $v^i$, and $s$,
by expanding $p(\rhoc + \delta \rho, s) = (\rhoc + \delta \rho)^2 \partial
e(\rhoc+\delta \rho, s)/\partial \rho$ around $\rho = \rhoc$. According to
relation \eqref{eq:rho}, the difference $\delta \rho \equiv \rho - \rhoc$
splits into \emph{(i)} a ``conservative'' term of order $1/c^2$, which we shall
note $\rho_{(2)\, {\rm cons}}$ henceforth, \emph{(ii)} a ``reaction''
term of order $1/c^7$, plus \emph{(iii)} contributions of order $1/c^4$,
$1/c^6$, and $\cO(1/c^8)$ that are irrelevant in our calculation. Thus,
$\cO([\delta \rho]^2) = 1/c^4 (\ldots) + 1/c^6 (\ldots) + \cO(1/c^8)$, and
the quadratic term of the Taylor series can be omitted.
\begin{align}
p(\rhoc + \delta \rho,s) =&~
p(\rhoc, s) + \delta \rho \frac{\partial p}{\partial \rho}(\rhoc, s) + 
\cO \left([\delta \rho]^2 \right) \nonumber \\
=&~ p(\rhoc, s) + \left(\rho_{(2)} + \rho_{(7)} \right) \frac{\partial
p}{\partial \rho}(\rhoc, s) \nonumber \\ & + \frac{1}{c^4} ( \ldots ) +
\frac{1}{c^6} ( 
\ldots ) +  \cO \left( \frac{1}{c^8} \right) \, . \label{eq:p_intermediate}
\end{align}
We define the coordinate pressure $\pc$ to be the pressure of a fictitious
fluid of scalar density $\rhoc$ and of entropy $s$ in our coordinate grid:
$\pc \equiv p(\rhoc, s)$. Its numerical value \emph{a priori} differs from
that of the actual pressure $p \equiv p(\rho(\rhoc, v^p, s) ,s)$. We introduce
in the same way a coordinate enthalpy $\hc$ as well as a coordinate adiabatic
index $\gammac = \partial \ln \pc/\partial \ln \rhoc$, etc. Whereas $h \neq
\hc$ and $\gamma \neq \gammac$ in general, $\gammac$ identifies with the usual
adiabatic index $\gamma = \partial \ln p/\partial \ln \rho$ when the equation
of state is assumed to be polytropic. Though $\gammac$ happens to be constant
in the latter case, it usually depends on time and on the field point:
$\gammac = \gammac(\mathbf{x}, t)$. Within
the preceding notation, equation \eqref{eq:p_intermediate} becomes:
\begin{align} \label{eq:p}
p =& \pc \left[1 - \frac{\gammac}{c^2} \left( 3 \Uc + \frac{1}{2} v^2 \right)
\right] \nonumber \\ & - \frac{\pc \gammac}{2} \left( [(g'_{ii})_{(7)}]_{(7)
    \, {\rm reac}} +  
[(g'_{ij})_{(5)}]_{(5) \, {\rm reac}} \frac{v^i v^j}{c^2} \right) \nonumber \\ 
&+ \frac{1}{c^4} ( \ldots ) + 
\frac{1}{c^6} ( \ldots ) + \cO \left( \frac{1}{c^8} \right) \, .
\end{align}
This entails notably that $h = \hc + ( \ldots )/c^2 + (\ldots )/c^4
+ \cO(1/c^6)$.
Finally, the $1$ PN $+~3.5$ PN mass density $\sigma = (T^{00} + T^{ii})/c^2$ is
derived by inserting expressions \eqref{eq:gmn}, \eqref{eq:u0},
\eqref{eq:rho}, and \eqref{eq:p} into the stress-energy tensor $T^{\mu\nu} =
\rhoc u^0 (1 + h/c^2 ) v^\mu v^\nu/\sqrt{-g'} + p g'^{\mu\nu}$. We arrive at
\begin{align} \label{eq:sigma} 
\sigma &= \rhoc + \frac{1}{c^2} \left[\rhoc c^2 \left(1 + u^0_{(2) \, {\rm
	  cons}} + 
u^0_{(7) \, {\rm reac}} \right) \right. \nonumber \\ & \qquad \; \left.
\times \left(1 - (\sqrt{-g'})_{(2) \, {\rm cons}}  -  (\sqrt{-g'})_{(7) \, {\rm
      reac}} \right) - \pc \right] 
\nonumber \\ &+ \frac{1}{c^2}  \left[\rhoc v^2 + \pc \left(\delta^{ii} -  
[(g'^{ii})_{(5)}]_{(5) \, {\rm reac}}\right) \right] \nonumber \\
=&~ \rhoc + \frac{\rhoc}{c^2} \left(\frac{3}{2} v^2 - \Uc + \hc + 2
\frac{\pc}{\rhoc}  \right) + \rhoc
\bigg([(g'_{00})_{(7)}]_{\genfrac{}{}{0pt}{1}{(7)}{\rm reac}} \nonumber \\ &
\quad\; -  \frac{1}{2} [(g'_{ii})_{(7)}]_{(7) \, {\rm reac}} 
+ \frac{1}{2} [(g'_{ij})_{(5)}]_{(5) \, {\rm reac}} \frac{v^i v^j}{c^2}
\bigg) \, ,
\end{align}
noticing that $[(g'^{ii})_{(5)}]_{(5) \, {\rm
reac}} = - [(g'_{ii})_{(5)}]_{(5) \, {\rm reac}} + \cO(1/c^7) = \cO(1/c^7)$. 
The Poisson integral of the above expansion gives the potential $U$ in the
$\{\rhoc, v^i, s\}$ representation, modulo an unimportant prefactor $G$.
\begin{align}
U =&~ \Uc + \frac{G}{c^2} \integ \frac{d^3\mathbf{y}}{|\mathbf{x} -
  \mathbf{y}|}~\rhoc \left(\frac{3}{2} v^2 - \Uc + \hc + 2
\frac{\pc}{\rhoc}  \right) \nonumber\\ &
+ G \integ \frac{d^3\mathbf{y}}{|\mathbf{x} - \mathbf{y}|} ~
\rhoc \bigg([(g'_{00})_{(7)}]_{\genfrac{}{}{0pt}{1}{(7)}{\rm reac}} -
\frac{1}{2} [(g'_{ii})_{(7)}]_{\genfrac{}{}{0pt}{1}{(7)}{\rm reac}} \nonumber\\
& \qquad \qquad \qquad \quad  +
\frac{1}{2} [(g'_{ij})_{(5)}]_{(5) \, {\rm reac}} \frac{v^i v^j}{c^2}
\bigg) \nonumber \\ & + \frac{1}{c^4} ( \ldots ) + \frac{1}{c^6} ( \ldots ) +
\cO \left( \frac{1}{c^8} \right) \, . \label{eq:U}
\end{align}
All other potentials can be cast into a similar form. The leading orders of
the conservative and reaction part are determined by that of $\sigma$ and
$\sigma_i$. As the fifth order corrections of both mass and current density
vanish, we have 
$\chi = \chi_* + \chi_{(2) \, {\rm cons}} + \chi_{(7) \, {\rm reac}}$ and
$U_i = \Uc_i + U_{i(2) \, {\rm cons}} + U_{i(7) \, {\rm reac}}$, modulo
irrelevant terms.

We are now in measure to isolate the ``reaction'' field 
$(g'_{\mu\nu})_{\rm reac}$. It essentially consists of the odd metric
\eqref{eq:gmn} to which add the reaction terms generated by the
post-Newtonian part of $(g'_{\mu\nu})_{\rm even} = g'_{\mu\nu} -
(g'_{\mu\nu})_{\rm odd}$ regarded as a function of the matter variables
$\sigma$ and 
$\sigma_i$. The latter terms do not contribute to $(g'_{0i})_{\rm reac}$ or
$(g'_{ij})_{\rm reac}$ at the $3.5$ PN order, but enter the
$00$ component of $(g'_{\mu\nu})_{\rm reac}$,
\begin{subequations} \label{eq:gmnreac}
\begin{align}
(g'_{00})_{\rm reac} &= (g'_{00})_{\rm odd} + \frac{2}{c^2} U_{(7) \, {\rm
reac}} + \cO \left(\frac{1}{c^{10}} \right) \, , \\
(g'_{0i})_{\rm reac} &= (g'_{0i})_{\rm odd} - \frac{1}{c^3} A_{i(7) \, {\rm
reac}} + \cO \left(\frac{1}{c^9} \right) \nonumber \\ & = (g'_{0i})_{\rm odd}
+  \cO \left(\frac{1}{c^9} \right) \, , \\
(g'_{ij})_{\rm reac} &= (g'_{ij})_{\rm odd} + \delta^{ij} 
\frac{2}{c^2} U_{(7) \, {\rm reac}} + \cO \left(\frac{1}{c^8} \right)
\nonumber \\ & =
(g'_{ij})_{\rm odd} + \cO \left(\frac{1}{c^8} \right) \, .
\end{align}
\end{subequations}
By virtue of equations \eqref{eq:gmn}, \eqref{eq:frechetP}, \eqref{eq:U}, and
\eqref{eq:gmnreac}, we finally obtain the $2.5$ and $3.5$ PN components of
the metric as
\begin{widetext}
\begin{subequations} \label{eq:gmn_explicit}
\begin{align}
(g'_{00})_{(9) \, {\rm reac}} =& \frac{4 G}{5c^9} \bigg( -  {I_2}_{kl}^{(3)}
x^k \partial_l 
\Uc + Q_{kl}^{(3)} x^k \left(- \partial_l U_2 + 2 \Uc \partial_l \Uc \right)
 + Q_{kl}^{(4)} x^k \left(- \frac{1}{2} x^l \partial_t \Uc
+ \Apot_l \right) + \frac{5}{126} Q_{klm}^{(5)} x^k x^l 
\partial_m \Uc \nonumber
\\ & + Q_{kl}^{(5)} x^k \left( \frac{17}{42} x^l x^m \partial_m \Uc -
\frac{11}{42} r^2 \partial_l \Uc \right) - \frac{8}{9} \epsilon_{klm}
S_{mn}^{(4)} x^l x^n \partial_k \Uc 
- 2 U G \integ \frac{d^3\mathbf{y}}{|\mathbf{x} - \mathbf{y}|}~ Q_{kl}^{(3)}
y^k \partial_l \rhoc \nonumber \\ & + G \integ
\frac{d^3\mathbf{y}}{|\mathbf{x} - \mathbf{y}|} \left[ {I_2}_{kl}^{(3)}
y^k \partial_l \rhoc + Q_{kl}^{(3)} y^k \left(\rhoc \partial_l \Uc +\partial_l
(\rhoc \delta) \right) -  3 \rhoc v^k v^l
Q_{kl}^{(3)}  - \frac{5}{126} Q_{klm}^{(5)} y^k y^l \partial_m \rhoc  \right.
\nonumber \\ & +
Q_{kl}^{(4)} y^k \left(\frac{1}{2} y^l \partial_t \rhoc - 
4 \rhoc v^l \right) + Q_{kl}^{(5)} y^k \left(-
\frac{17}{42} y^l y^m \partial_m \rhoc + \frac{11}{42} |\mathbf{y}|^2
\partial_l \rhoc - \rhoc y^l \right) - \frac{8}{9} \epsilon_{klm} S_{mn}^{(4)}
y^l y^n \partial_k \rhoc  \nonumber \\ & \left. -  G \rhoc \integ
  \frac{d^3\mathbf{y}'}{|\mathbf{y} - \mathbf{y}'|} \left(Q_{kl}^{(3)}
y^k \partial_l \rhoc \right)[\mathbf{y}'] \right] \bigg) \, , \\
(g'_{0i})_{(8) \, {\rm reac}} =& - \frac{8 G}{5c^8} \bigg( - 
\frac{1}{4} Q_{kl}^{(3)} x^k \partial_l \Apot_i - \frac{1}{4} Q_{ik}^{(3)}
\Apot_k 
+  Q_{ik}^{(4)} x^k \Uc  + \frac{G}{8} \partial_i
\partial_t  \integ d^3\mathbf{y}~|\mathbf{x} - \mathbf{y}| \left[
 Q_{kl}^{(3)} y^k \partial_l \rhoc \right] \nonumber \\ &  + 
G \integ \frac{d^3\mathbf{y}}{|\mathbf{x} - \mathbf{y}|} \left[
Q_{kl}^{(3)} y^k \partial_l (\rhoc v^i) -
\rhoc v^k Q_{ik}^{(3)} - \rhoc y^k Q_{ik}^{(4)} \right]
 \bigg) \, , \label{eq:g00_explicit} \\
(g'_{ij})_{(7) \, {\rm reac}} =& \frac{4 G}{5c^7} \bigg( -
{I_2}^{(3)}_{ij} - 2 Q_{ij}^{(3)} \Uc + 
\frac{5}{63} Q_{ijk}^{(5)} x^k + \frac{2}{7} x^k x^{(i} Q_{j)k}^{(5)} -
\frac{11}{42} Q_{ij}^{(5)} r^2 - 
\frac{8}{9} \epsilon_{kl(i} S_{j)l}^{(4)} x^k \bigg) \nonumber \\ &
 +  \frac{4 G}{5}\delta^{ij} \bigg(- \frac{2}{21} x^k x^l Q_{kl}^{(5)} -
Q^{(3)}_{kl} x^k \partial_l \Uc  +  
G \integ \frac{d^3\mathbf{y}}{|\mathbf{x} - \mathbf{y}|}~Q^{(3)}_{kl} y^k
\partial_l \rhoc \bigg) \, ,
\end{align}
\end{subequations}
\end{widetext}
with $\Apot_i \equiv 4 \Uc_i + \partial_i \partial_t \chi_*/2$, $U_2 \equiv
c^2 U_{(2) \, {\rm cons}}$, ${I_2}_{ij} \equiv c^2 (I_{ij})_{(2)
\, {\rm cons}}$, and $S_{ij} \equiv J_{ij(0) \, {\rm cons}} = \integ
d^3\mathbf{y}~ \epsilon_{km<i} \hat{y}^{j>k} \rhoc v^m$.

\section*{$\mathbf{3.5}$ PN reaction force} \label{sec:reaction}
The evolution equations including the $3.5$ PN gravitational damping are
deduced from the conservation of the stress-energy tensor
\begin{equation} \label{eq:conservation_Tmn}
 \partial_\alpha \left(\sqrt{-g'} T^\alpha_{\, \mu} \right) =
\frac{1}{2}  \sqrt{-g'} T^{\alpha\beta} \partial_\mu g'_{\alpha\beta}\, ,
\end{equation}
and from that of the entropy
\begin{equation} \label{eq:conservation_s}
\partial_t s + v^i \partial_i s =0 \, .
\end{equation}
The projection of formula \eqref{eq:conservation_Tmn} on $u^\mu$,
together with formula \eqref{eq:conservation_s}, leads to the
continuity equation
\begin{equation} \label{eq:conservation_rhoc}
\partial_t \rhoc + \partial_i (\rhoc v^i) = 0 \, ,
\end{equation}
whereas its space part is equivalent to the Euler equation
\begin{multline} \label{eq:euler}
\partial_j \left[\rhoc \left(1 + \frac{h}{c^2} \right) u_i v^j c\right] +
\partial_t \left[\rhoc \left(1 + \frac{h}{c^2}\right) u_i c\right] \\ + 
\partial_i (\sqrt{-g'} p)
= \frac{1}{2} \sqrt{-g'} T^{\alpha\beta} \partial_i g'_{\alpha\beta} \, .
\end{multline}
In the absence of coupling between radiative and non-radiative effects, the
$1$ PN + $3.5$ PN dynamics of the adiabatic fluid subjected to 
gravitational damping is described by the post-Newtonian expansion of
equalities \eqref{eq:conservation_s},
\eqref{eq:conservation_rhoc}, \eqref{eq:euler}, including the Newtonian, 1
PN, $2.5$ PN, as well as the $3.5$ PN order for fixed $\rhoc$, $v^i$, and $s$.

We first identify the momentum density of the
fluid\footnote{Note that the momentum density $M_i$ is referred to as $\pi_i$
in paper \cite{BDS90}} $M_i \equiv \rhoc (1 + h/c^2) u_i c = \sqrt{-g'}
T^0_{\, i}/c$ on the left-hand side of \eqref{eq:euler}. The 
space part $u_i$ of the covariant velocity is related to the variables
$\rhoc$, $v^i$, and $s$ through the equality $u_i = v^i/c + [u_0 (g'_{i0} +
g'_{ij} v^j/c)]_{(3+6+8)} + ( \ldots )/c^5 + ( \ldots )/c^7 + \cO(1/c^9)$.
After replacing $u^0$ by its value \eqref{eq:u0} in the definition of $M_i$
and expanding up to the $1$ PN + $3.5$ PN order with the
help of equations \eqref{eq:gmn}, we arrive at
\begin{align} \label{eq:Mi}
M_i =& \rhoc v^i + \frac{\rhoc}{c^2} \left(\hc v^i + 3 \Uc v^i - \Apot_i +
\frac{1}{2} v^2 v^i \right) \nonumber \\ &  + \rhoc v^j (g'_{ij})_{(5+7) \,
{\rm reac}} \nonumber \\ &+ \frac{1}{c^2} \rhoc
\left(\hc + \frac{1}{2} v^2 + \Uc \right) v^j (g'_{ij})_{(5) \, {\rm reac}}
\nonumber \\ & + \rhoc c (g'_{0i})_{(8) \, {\rm reac}} \nonumber \\ & +
\frac{1}{2} \rhoc v^i \left((g'_{00})_{(7) \, {\rm reac}} + (g'_{kl})_{(5) \,
    {\rm reac}} \frac{v^k v^l}{c^2} \right) \nonumber \\ & + \frac{1}{c^4} (
\ldots ) + \frac{1}{c^6} ( \ldots ) + \cO \left(\frac{1}{c^8} \right) \, .
\end{align}
This expression contains the term $\rhoc (g'_{ij})_{(5+7)} v^j$ involving in
particular the third order time derivative of the 1 PN quadrupole $I_{ij}$ and
the fifth time derivative of the Newtonian mass multipole moments $Q_{ij}$ and
$Q_{ijk}$. Therefore, the extra time derivation appearing on the left-hand
side of the Euler equation \eqref{eq:euler} induces terms in $I_{ij
(2) \, {\rm cons}}^{(4)}$, $Q_{ij}^{(6)}$, and $Q_{ijk}^{(6)}$, which make us
lose all benefit from the gauge transformation. In order to cure
that, we modify our system of variables according to paper \cite{BDS90}.
We keep the baryonic density as well as the entropy, but substitute to the
coordinate velocity $v^i$ the more appropriate ``momentum velocity'' $w_i
\equiv M_i/\rhoc$. The relation between $w_i$ and $v^i$ is found by inverting
equation \eqref{eq:Mi} iteratively. At the leading order, we have $M_i = \rhoc
v^i + \cO (1/c^2)$, hence $w_i = v^i + \cO(1/c^2)$. This result is used to
compute the 1 PN correction of $v^i$ as a function of 
$w_i$: $M_i = \rhoc v^i - \rhoc (- \hc w_i - 3 \Uc w_i + \Apot_i - w^2
w_i/2)/c^2 + \cO(1/c^4)$, hence $v^i$, which is in its turn
inserted into equation \eqref{eq:Mi} truncated at the next approximation, etc.
We get in the end:
\begin{align} \label{eq:vi}
v^i &= w_i + \frac{1}{c^2} \left(- \hc w_i - 3 \Uc w_i + \Apot_i  -
  \frac{1}{2} w^2 w_i \right)  \nonumber \\ & - w_j (g'_{ij})_{(5) \, {\rm
  reac}} - w_j (g'_{ij})_{(7) \, {\rm reac}} \nonumber \\ & + \frac{1}{c^2}
\left(5 \Uc + \frac{1}{2}
w^2 + \hc \right) w_j (g'_{ij})_{\genfrac{}{}{0pt}{1}{(5)}{\rm cons}} -
  \frac{\Apot_j}{c^2}  
(g'_{ij})_{\genfrac{}{}{0pt}{1}{(5)}{\rm reac}} \nonumber \\
& - c (g'_{0i})_{\genfrac{}{}{0pt}{1}{(8)}{\rm reac}} + \frac{w_i}{2} \left(-
  (g'_{00})_{\genfrac{}{}{0pt}{1}{(7)}{\rm reac}} +
  (g'_{kl})_{\genfrac{}{}{0pt}{1}{(5)}{\rm reac}} \frac{w_k 
  w_l}{c^2} \right) \nonumber 
\\ & + \frac{1}{c^4} ( \ldots ) + \frac{1}{c^6} ( \ldots ) + \cO
\left(\frac{1}{c^8} \right) \, .
\end{align}

Up to now, we have reduced the kinetic part of the Euler equation
\eqref{eq:euler} to 
$\partial_j (\rhoc w_i v^j) + \partial_t (\rhoc w_i)$, or
equivalently $\rhoc (v^j \partial_j w_i + \partial_t w_i)$, after
the continuity equation. The pressure force $F_i^{\rm
press}\equiv -\partial_i (\sqrt{-g'} p)/\rhoc$, as for it, is 
derived from the post-Newtonian expansion \eqref{eq:p} and
\eqref{eq:volume} of $p$ and $\sqrt{-g'}$ respectively, supplemented by the
two approximate 
equalities $U = \Uc + (\ldots)/c^2 + (\ldots)/c^4 + \cO(1/c^6)$,
$(g'_{ii})_{(7)} = (g'_{ii})_{(7) \, {\rm reac}}$, and $v^i = w_i - w_j
(g'_{ij})_{(5) \, {\rm reac}} +$ irrelevant terms.
\begin{widetext}
\begin{align}
F_i^{\rm press} =& - \frac{1}{\rhoc} \partial_i \bigg\{\pc +
\frac{1}{c^2} \left[2 \Uc - \gammac \left(3 \Uc + \frac{1}{2} v^2 \right)
\right] \pc \nonumber  \\ & \qquad \quad \; + \frac{\pc}{2}
\bigg[(g'_{jj})_{\genfrac{}{}{0pt}{1}{(7)}{\rm reac}} - (g'_{00})_{(7) \, {\rm
    reac}}  - \gammac \left((g'_{jj})_{\genfrac{}{}{0pt}{1}{(7)}{\rm reac}} +
  (g'_{jk})_{\genfrac{}{}{0pt}{1}{(5)}{\rm reac}} \frac{v^j 
v^k}{c^2} \right) \bigg] \bigg\} + \frac{1}{c^4} (\ldots) +
\frac{1}{c^6} (\ldots) + \cO \left(\frac{1}{c^8} \right) \nonumber \\ =& -
\bigg[1 + \frac{1}{c^2} \left(2 \Uc - \gammac \left(3 
\Uc + \frac{1}{2} w^2\right)  \right) +
\frac{1}{2} \left( (g'_{jj})_{\genfrac{}{}{0pt}{1}{(7)}{\rm reac}} -
  (g'_{00})_{\genfrac{}{}{0pt}{1}{(7)}{\rm reac}} \right)   
- \frac{\gammac}{2} \left((g'_{jj})_{\genfrac{}{}{0pt}{1}{(7)}{\rm reac}} -
  (g'_{jk})_{\genfrac{}{}{0pt}{1}{(5)}{\rm reac}} 
\frac{w_j w_k}{c^2} \right)\bigg] \frac{\partial_i
\pc}{\rhoc} \nonumber \\ & \qquad \; - \frac{\pc}{\rhoc} \partial_i 
\bigg[\frac{1}{c^2} \left(2 \Uc - \gammac \left(3
\Uc + \frac{1}{2} w^2\right)  \right) + \frac{1}{2}
\left( (g'_{jj})_{(7) \, 
{\rm reac}} - (g'_{00})_{(7) \, {\rm reac}} \right) \nonumber \\ & \qquad \;
- \frac{\gammac}{2} \left((g'_{jj})_{(7) \, {\rm reac}} - (g'_{jk})_{(5) \,
{\rm reac}} 
\frac{w_j w_k}{c^2} \right) \bigg] + \frac{1}{c^4} (\ldots
) + \frac{1}{c^6} (\ldots) + \cO \left(\frac{1}{c^8} \right)\, .
\end{align}
While both pressure and density tend toward zero in the region of space-time
extending outside the system, the ratio $\partial_i \pc/\rhoc$ remains finite
by virtue of
the thermodynamical relation $\partial_i \pc/\rhoc = \partial_i \hc - T_*
\partial_i s$, with $T_* = \partial e(\rhoc, s)/\partial s$ representing the
coordinate temperature. The last term of the Euler equation still
to be evaluated is the gravitational force $F_i^{\rm grav} \equiv \partial_i
g'_{\alpha \beta} \sqrt{-g'} T^{\alpha \beta}/(2\rhoc)$. The components
$g'_{\mu\nu}$ of the metric in the new gauge and the square root of minus the 
corresponding determinant are given by equations \eqref{eq:gmn_explicit}
and \eqref{eq:volume} respectively. The stress-energy tensor is obtained from
the mass density \eqref{eq:sigma}, the current density $\sigma_i = \rhoc v^i +
\cO (1/c^2)$, as well as the stress density \eqref{eq:sigmaijreac}, by
inverting the identities $\sigma = (T^{00} + T^{ii})/c^2$, $\sigma_i =
T^{0i}/c$, and $\sigma_{ij} = T^{ij}$. All contributions of order
$1/c^4$, $1/c^6$, and $\cO(1/c^8)$ are neglected as usual. We find:
\begin{align}
F_i^{\rm grav} &= \frac{1}{\rhoc} \bigg\{ 
\frac{1}{2} \partial_i \left[-1 + \frac{2}{c^2} \Uc - \frac{2}{c^4} \Uc^{\! \!
	2} \right] \left[\sqrt{-g'} (c^2 \sigma -
\sigma_{jj})\right]_{\genfrac{}{}{0pt}{1}{(-2+0)}{\rm cons}}
+ \partial_i \left[-\frac{\Apot_j}{c^3} \right] \left[\sqrt{-g'} c
\sigma_j \right]_{\genfrac{}{}{0pt}{1}{(0)}{\rm cons}}  \nonumber \\ & \qquad
\; + \frac{1}{2} \partial_i
\left[\delta^{jk} \left(1 + \frac{2}{c^2} \Uc \right) \right] \left[\sqrt{-g'}
\sigma_{jk}\right]_{\genfrac{}{}{0pt}{1}{(0)}{\rm cons}}
+ \frac{1}{2} \partial_i \left[-1 + \frac{2}{c^2} \Uc
\right] \left[\sqrt{-g'} (c^2 \sigma -
\sigma_{jj})\right]_{\genfrac{}{}{0pt}{1}{(5)}{{\rm reac}}} \nonumber \\ &
\qquad + \partial_i \left[-\frac{\Apot_j}{c^3} \right] \left[\sqrt{-g'} c
  \sigma_j \right]_{\genfrac{}{}{0pt}{1}{(4)}{\rm reac}} + \frac{1}{2}
\partial_i \left[\delta^{jk} \left(1 + \frac{2}{c^2} \Uc \right) \right]
\left[\sqrt{-g'} \sigma_{jk}\right]_{\genfrac{}{}{0pt}{1}{(5)}{\rm reac}}
\nonumber \\ & \qquad +
\frac{1}{2} \partial_i (g'_{00})_{\genfrac{}{}{0pt}{1}{(7+9)}{\rm reac}} \!
\left[\sqrt{-g'} (c^2 \sigma -
  \sigma_{jj})\right]_{\genfrac{}{}{0pt}{1}{(-2+0)}{\rm cons}}
+ \partial_i (g'_{0j})_{\genfrac{}{}{0pt}{1}{(6+8)}{\rm reac}} \!
\left[\sqrt{-g'} c \sigma_{j}\right]_{(-1) \, {\rm cons}} \nonumber \\ &
\qquad + \frac{1}{2} \partial_i (g'_{jk})_{\genfrac{}{}{0pt}{1}{(5+7)}{\rm
    reac}} \left[\sqrt{-g'} \sigma_{jk} \right]_{(0+2) \, {\rm cons}}
+ \frac{1}{c^4} ( \ldots) + \frac{1}{c^6} ( \ldots ) +
\cO\left( \frac{1}{c^8} \right)  \bigg\} \nonumber \\
& = \partial_i \Uc \left[1 + \frac{1}{c^2}\left(\frac{3}{2} v^2 - \Uc + \hc + 2
\frac{\pc}{\rhoc} \right) \right] + \frac{1}{c^2} \partial_i U_2 -
\frac{1}{c^2} v^j \partial_i \Apot_j \nonumber \\ & \qquad + \frac{1}{2} c^2
\partial_i (g'_{00})_{(7) \, {\rm reac}} + 
\frac{1}{2} \partial_i \Uc \left((g'_{00})_{(7) \, {\rm reac}} + 
(g'_{jk})_{(5) \, {\rm reac}} \frac{v^j v^k}{c^2} \right) \nonumber \\ &
\qquad + \frac{1}{2} \partial_i (g'_{00})_{(7) \, {\rm reac}} \left(\hc + \Uc
  + \frac{1}{2} v^2 - \frac{\pc}{\rhoc} \right) + 
\frac{c^2}{2} \partial_i (g'_{00})_{(9) \, {\rm reac}} + c v^j\partial_i
(g'_{0j})_{(8) \, {\rm reac}} \nonumber \\ & \qquad  + \frac{1}{2} \left(v^j 
v^k \partial_i (g'_{jk})_{(7) \, {\rm reac}} + \frac{\pc}{\rhoc} \partial_i
(g'_{jj})_{(7) \, {\rm reac}} \right) + \frac{1}{c^4} ( \ldots) + \frac{1}{c^6}
( \ldots) + \cO \left( \frac{1}{c^8} \right) \, . \label{eq:Figrav}
\end{align}
\end{widetext}

The indetermination of the ratio $\pc/\rhoc$ for $\pc$ and $\rhoc$ both going
to zero is raised by use of the relation $\pc/\rhoc = \rhoc \partial e(\rhoc,
s)/\partial \rhoc$. The replacement of $v^i$ in $F^{\rm grav}_i$
according to equation \eqref{eq:vi} amounts to \emph{(i)} changing $v^i$ into
$w_i$ at each of its occurrences, and \emph{(ii)} adding the $3.5$ PN term:
$\delta F_i^{\rm grav} = -3 \partial_i \Uc  w_j w_k (g'_{jk})_{(5) \,
{\rm reac}}/c^2 + \partial_i \Apot_j w_k (g'_{jk})_{(5) \, {\rm reac}}/c^2$ to
the resulting expression.
Knowing $v^i$, $F^{\rm grav}_i$, and $F^{\rm press}_i$, we write the Euler
equation as:
\begin{equation} \label{eq:euler_explicit}
\partial_t w_i + v^j \partial_j w_i = F^{\rm grav}_i + F^{\rm press}_i \, .
\end{equation}

The evolution equations for the fluid are now established. They are
parametrized by the Newtonian potential $\Uc$, the post-Newtonian potential
$\Apot_i = 4 \Uc_i + \partial_i \partial_t \chi_*/2$, as well as the $2.5$ and
$3.5$ PN reaction metric components $(g'_{00})_{(7) \, {\rm reac}}$,
$(g'_{00})_{(9) \, {\rm reac}}$, $(g'_{0i})_{(8) \, {\rm reac}}$,
$(g'_{ij})_{(5) \, {\rm reac}}$, and $(g'_{ij})_{(7) \, {\rm reac}}$. Formulas
\eqref{eq:gmn2.5PN} and \eqref{eq:gmn_explicit} show that these components
depend themselves on $\Uc$, $\Apot_i$, and on the Poisson integrals coming
from the functional derivation. Among these various quantities, the only super
potential, $\chi_*$, appears exclusively through its gradient, $\partial_i
\chi_* = G \integ d^3\mathbf{y}~\rhoc (x^i - y^i)/|\mathbf{x}- \mathbf{y}|$,
which can be put under the form $\partial_i \chi_* = x^i \Delta^{-1} (-4 \pi G
\rhoc ) -\Delta^{-1} (-4 \pi G x^i \rhoc)$. Consequently, all the elementary
field quantities employed in the present formulation are solutions of
decoupled Poisson equations with compact support sources. This simple elliptic
structure fits particularly well to numerical computations. The only
complication, apart from that due to the mere length of the expressions, comes
from the double integration arising in $(g'_{00})_{(9) \, {\rm reac}}$.
Indeed, the source terms $\rhoc Q_{kl}^{(3)} y^k \partial_l \Uc$ and $\rhoc
\Delta^{-1} \left(-4 \pi G Q_{kl}^{(3)} y^k \partial_l \rhoc \right)$ present
in the second member of equation \eqref{eq:g00_explicit} are basically
themselves Poisson potentials. This nonlinear effect originates in the fact
that the integrand of $U_{(7) \, {\rm reac}}$ involves certain pieces of the
$2.5$ and $3.5$ PN gravitational field, namely $(g'_{00})_{(7) \, {\rm
    reac}}$, $(g'_{ii})_{(7) \, {\rm reac}}$, and $(g'_{ij})_{(5) \, {\rm
    reac}}$. It occurs either in the original Burke-Thorne-like grid or in the
coordinate system defined in section \ref{sec:gauge}, as long as we confine
ourselves to the $\{\rhoc,w_i,s\}$ representation, but can be cured by
redefining the density variable as $\tilde{\rho} = \rhoc + \sigma_{(7)}$. This
results in a slight modification of the present formalism. The corresponding
set of equations is presented in Appendix \ref{sec:appA}.

Our last task in the achievement of our initial goal will consist in
reducing the number of temporal derivatives remaining in the
metric \eqref{eq:gmn_explicit} to one. This will be done by eliminating
$\partial_t \rhoc$, $\partial_t w_i$, and $\partial_t s$ with the help of the
entropy conservation, of the continuity equation, and of the Euler equation at
the Newtonian or post-Newtonian order. The multiple time derivatives we shall
have to deal with 
apply on the multipole moments. They are of the type $(d/dt)^n \integ
\dx~\rhoc(\mathbf{x}, t) f(\mathbf{x},t)$, with $n \in \{2,3, \ldots\}$. The
function $f$ depends on time through the evolution variables
$\rhoc$, $w_i$, and $s$:  $f(\mathbf{x},t) = f(\mathbf{x},\rhoc(\mathbf{y},t),
w_p(\mathbf{y},t), s(\mathbf{y},t))$. Each differentiation is performed
by means of the formula:
\begin{align} \label{eq:chandra}
  &\frac{d}{dt} \integ \dx~\rhoc(\mathbf{x},t) f(\mathbf{x},t) =
  \integ \dx~\partial_t \left[\rhoc(\mathbf{x},t) f(\mathbf{x},t)\right]
  \nonumber \\ 
  &= \integ \dx~\left[- f(\mathbf{x},t) \partial_i (\rhoc(\mathbf{x},t)
    v^i({\bf x},t)) + \rhoc(\mathbf{x},t) \partial_t f(\mathbf{x},t)\right]
  \nonumber \\ & = \integ \dx~\rhoc(\mathbf{x},t) \frac{df}{dt}(\mathbf{x},t)
  \, ,
\end{align}
where the last equality is obtained by integration by part. We recall that
the total derivative $df/dt = \partial_t f+ v^i \partial_i f$ represents the
temporal variation of the field $f$ for a fixed particle of fluid. It
satisfies the chain rule $df/dt =(d\rhoc/dt) \partial f/\partial
\rhoc + (dw_i/dt) \partial f/\partial w_i + (ds/dt) \partial f/\partial s$
for any $f=f(\rhoc,w_i,s)$. Taking advantage of the Euler equation
\eqref{eq:euler_explicit}, $dw_i/dt = F_i^{\rm grav} + F_i^{\rm press}$, as
well as the local
conservation of mass \eqref{eq:conservation_rhoc}, $d\rhoc/dt = - \partial_i
(\rhoc v^i) + v^i \partial_i \rhoc = -\rhoc \partial_i v^i$, and entropy
\eqref{eq:conservation_s}, $ds/dt = 0$, we get:
\begin{equation} \label{eq:fdot}
\frac{df}{dt} = - \frac{\partial f}{\partial \rhoc} \rhoc \partial_i v^i +
\frac{\partial f}{\partial w_i} (F^{\rm grav}_i + F^{\rm press}_i) \, .
\end{equation} 
This shows that the time derivative of a term $\integ \dx~\rhoc(\mathbf{x},t)
f(\mathbf{x},t)$ is itself of the type $\integ \dx~\rhoc(\mathbf{x},t)
g(\mathbf{x},t)$ with $g=g(\mathbf{x},\rhoc(\mathbf{y},t), w_p(\mathbf{y},t),
s(\mathbf{y},t))$. In this way, the number of time derivatives acting on the
initial integral can be brought down to $n-1$, $n-2$, etc. After $n-1$
iteration, we are left with a unique time derivative as required. The price to
pay is an increase of the expression lengths. The evaluation of $I_{ij}^{(3)}$
is particularly tedious since the mass quadrupole $I_{ij}$ is needed at the
order $1/c^2$. After specializing the definition \eqref{eq:moments} of
$I_{ij}$ to the desired level of approximation, we have
\begin{align*}
I_{ij} =& \integ \dx~\hat{x}^{ij} \rhoc \left[1 + \frac{1}{c^2}
\left(\frac{3}{2} w^2 - \Uc + \hc + 2 \frac{\pc}{\rhoc}  \right) \right]
\\ &~ + \frac{1}{14c^2} \frac{d^2}{dt^2} \integ \dx~ r^2 \hat{x}^{ij} \rhoc
- \frac{20}{21c^2} \frac{d}{dt} \integ \dx~ \hat{x}^{ijk} \rhoc w_k \\ &~ +
\cO \left(\frac{1}{c^3} \right) \, . 
\end{align*}
We then perform the time derivation along the procedure
explained above. The force per mass unit $F^{\rm grav}_i + F^{\rm press}_i$ is
limited to the Newtonian order:
$dw_i/dt = \partial_i \Uc - \partial_i \pc/\rhoc$. We are led to \cite{BDS90}
\begin{subequations}
\begin{equation} \label{eq:Iij}
I_{ij}  = E_{ij} + F_{ij} + G_{ij} + \cO\left(\frac{1}{c^3} \right) \, ,
\end{equation}
with
\begin{align}
E_{ij} & =  \textrm{STF} \integ \dx~ \rhoc x^i x^j \bigg[1
 +\frac{1}{c^2} \bigg(\frac{1}{2} 
w^2 - \Uc \nonumber \\ & \qquad \qquad \qquad \qquad \qquad \qquad \; + \hc
 -  \frac{\pc}{\rhoc}  \bigg)  \bigg] \, , \\ 
F_{ij} & = \textrm{STF} \integ \dx~ \frac{\rhoc}{c^2} \bigg[\frac{11}{21} r^2
w_i w_j - \frac{4}{7} x^i x^k w_k w_j \nonumber \\ & \qquad \qquad \qquad
 \qquad \qquad \qquad \; + \frac{4}{21} x^i x^j w^2 \bigg]\, ,
\end{align}
and
\begin{equation}
G_{ij} = \textrm{STF} \integ \dx ~\frac{\rhoc}{c^2} \left[
\frac{11}{21} r^2 x^i \partial_j \Uc - \frac{17}{21} x^i x^j x^k \partial_k
\Uc \right] \, .
\end{equation}
\end{subequations}
The computation of $I_{ij}^{(2)}$ from equation \eqref{eq:Iij} is long, but
straightforward. We make systematical use of formulas \eqref{eq:chandra} and
\eqref{eq:fdot}. The quantities $v^i$ and 
$F^{\rm grav}_i + F^{\rm press}_i$ are truncated at the 1 PN order in the
first source term of $E_{ij}$, at the Newtonian order in the rest. 
\begin{align*}
v^i =& w_i \left[1 - \frac{1}{c^2} \left(
\frac{1}{2} w^2 + \hc + 3 \Uc \right) \right] + \frac{1}{c^2} \Apot_i \, ,\\
\frac{dw_i}{dt} =& \partial_i \Uc - \frac{\partial_i
\pc}{\rhoc} + \frac{1}{c^2} \left(\frac{3}{2} w^2 + \hc + 2
\frac{\pc}{\rhoc} - \Uc \right) \partial_i \Uc \nonumber \\ & + \frac{1}{c^2}
\partial_i U_2 - \frac{1}{c^2} w_j \partial_i \Apot_j \\ & 
- \frac{1}{c^2}\partial_i \left[ \left(2 \Uc -
\gammac \left( \frac{1}{2} w^2 + 3 \Uc\right) \right) 
\frac{\pc}{\rhoc} \right] \, . 
\end{align*}
The total time derivative of the enthalpy $\hc$ is deduced from relation
\eqref{eq:fdot}: $d\hc/dt =  (d\rhoc/dt) (\partial \hc/\partial \pc) (\partial
\pc/\partial \rhoc)  = - \gammac \pc \partial_i w_i/\rhoc + \cO(1/c^2)$.
For treating the potentials $\Uc$, $\Apot_i$, etc., the operator
$d/dt$ is first put under its explicit form,
$d/dt= \partial_t + v^i \partial_i$, and the partial derivative $\partial_t$
is next applied on the source $S$ through the 
integral symbol. The source variation $\partial_t S$ is transformed by means
of the evolution equations at the Newtonian order. For convenience, we
may redefine $\Uc_i$ as $\Delta^{-1} (-4\pi G \rhoc w_i)$. This yields for
instance 
\begin{align*}
\frac{d}{dt} \Uc =&~
\Delta^{-1} \partial_t (-4 \pi G \rhoc) + v^i \partial_i \Uc \\ =&
- \Delta^{-1} [-4 \pi G \partial_i (\rhoc w_i)] + w_i \partial_i \Uc +
\cO\left(\frac{1}{c^2} \right) \\
=& - \partial_i \Uc_i + w_i \partial_i \Uc +
\cO\left(\frac{1}{c^2} \right) \, , \\
\frac{d}{dt} \Uc_i =&~ \Delta^{-1} \partial_t (-4 \pi G \rhoc w_i) + v^j
\partial_j \Uc_i \\ =& \Delta^{-1} [-4\pi G (-\partial_{j} (\rhoc w_i w_j) +
\rhoc \partial_i \Uc - \partial_i \pc)] \\ & \nonumber + w_j
\partial_j \Uc_i + \cO \left(\frac{1}{c^2} \right) \, . 
\end{align*}
As the potential $\partial_t (\partial_i \chi_*)$ occurs in
$I_{ij}^{(2)}$, its time derivative must be 
eliminated as well. In order not to increase the number 
of Poisson equations, we systematically replace its old expression by the new
one. It must be determined at the $2.5$ PN order, since $\Apot_i$ appears in
the 1 PN part of the gravitational force \eqref{eq:Figrav}.
\begin{align*}
\partial_t \partial_i \chi_* =&~
\partial_t \left[x^i \Delta^{-1} (-4 \pi G \rhoc ) - \Delta^{-1} (-4 \pi G
x^i \rhoc ) \right] \\
=& - x^i \partial_j  \Delta^{-1} (-4 \pi G \rhoc v^j) \\ & + \Delta^{-1} [-4
\pi G x^i \partial_j (\rhoc v^j)]\\
=& - x^i \left(\delta^{jk} - (g'_{jk})_{(5) \, {\rm reac}} \right) \partial_j
\Uc_k \\ & + \Delta^{-1} \left[-4 \pi G \left(\delta^{jk} -
(g'_{jk})_{(5) \, {\rm reac}} \right) x^i \partial_j (\rhoc w_k)\right] \\ &+
\frac{1}{c^2}  (\ldots) + \frac{1}{c^4} (\ldots) +  \cO
\left(\frac{1}{c^6}\right) 
\end{align*}
(the last form is obtained by using the Newtonian plus $2.5$ PN
approximation for the coordinate 
velocity \eqref{eq:vi}: $v^i = w_i - w_j (g'_{ij})_{(5) \, {\rm reac}}$).
It is sufficient to determine the total time derivative of $\partial_t
\partial_i \chi_*$ contributing to $I_{ij}^{(2)}$ at the Newtonian order. It
reads
\begin{align*}
\partial_t (\partial_t & \partial_i \chi_* )  \\  =
&\partial_t \left[-x^i \partial_j \Uc_j + \Delta^{-1} (-4\pi G x^i \partial_j
(\rhoc w_j) ) \right] \\ =& 
-x^i \partial_j \Delta^{-1} [-4\pi G (- \partial_k (\rhoc w_j w_k) \\ &
  \qquad \qquad \qquad \qquad +
\rhoc \partial_j \Uc - \partial_j \pc) ] \\ &+ \Delta^{-1} [-4\pi G x^i 
(-\partial_{jk} (\rhoc w_j w_k) \\ & \qquad \qquad \qquad \quad \, + 
\partial_j (\rhoc \partial_j \Uc) - \Delta \pc) ] \, . 
\end{align*}
After ruling out all temporal differentiation from $I_{ij}^{(2)}$, our result
is finalized by removing, whenever possible, the space derivatives of the
baryonic density $\rhoc$ and the pressure $\pc$ by integration by
part. Remark that the space and time derivatives of
$\gammac$ exactly cancel.
All quantities $S_{ij}^{(n)}$ and $Q_{ijk}^{(n)} = (d/dt)^n \integ
\dx~\hat{x}^{ijk} \rhoc$ for $n \le 4$ are rewritten at Newtonian level
following the preceding method. 

Now, $S_{ij}$ and $Q_{ijk}$, as well as the hexadecapole $Q_{ijkl} = \integ
\dx~\hat{x}^{ijkl} \rhoc$, can be formally assimilated to the multipoles that
parametrize the wave form $h_{ij}^{\rm rad}$ in some ``radiative coordinate''
grid $X^\mu=(T,\mathbf{X})$ up to the $3.5$ order, as explained in paper
\cite{BDS90}. After actually achieving the substitution into the wave form,
the new expression for $h_{ij}^{\rm rad}(\mathbf{X},T)$ involves one time
derivation only, and is thus suitable for numerical computations. In
conclusion, the present formalism allows one to build an accurate $1$ PN +
$3.5$ PN code (\emph{i}) that models the fluid dynamics including the $3.5$ PN
damping effect, and (\emph{ii}) that permits gravitational wave extraction.
All the field equations are of Poisson type, and have compact support sources.
The number of time derivatives is never higher than one.

Let us list in the end the quantities to be computed and the equations to be
solved for modeling the evolution of an adiabatic fluid at the $1$ PN +
$3.5$ PN approximation.
For reader's convenience, we follow
the presentation of Blanchet-Damour-Sch\"afer \cite{BDS90}. The
notations are as close to theirs as possible, though not
identical.\footnote{We have the following correspondences:\\
\hspace*{0.5cm} \begin{tabular}{|l|c|c|c|c|} \hline
our notation & $\rhoc$ & $e$ & $\Uc_i$ & $\Apot_i$ \\ \hline
notation of paper \cite{BDS90} & $r_*$ & $e/r_*$ & $U_i$ & $A_i$ \\
\hline
 \end{tabular}}
It is useful to pose:
\begin{subequations} \label{eq:FS_equations}
\begin{flalign}
&U_5 = \frac{c^7}{2} (g'_{00})_{(7) \, {\rm reac}} \, , & \\
&U_7 = \frac{c^9}{2} (g'_{00})_{(9) \, {\rm reac}} + 3 c^5 \Delta^{-1}
 \left(-4 \pi G \rhoc v^i_{(5) \, {\rm reac}} w_i \right) \, , & \\
&{U_5}_i = - \frac{c^8}{4} (g'_{0i})_{(8) \, {\rm reac}} \, , & \\
&{h_5}_{ij} = \frac{c^5}{2} (g'_{ij})_{(5) \, {\rm reac}} \, , & \\
&{h_7}_{ij} = \frac{c^7}{2} (g'_{ij})_{(7) \, {\rm reac}} - \delta^{ij} U_5
\,. 
\end{flalign}
The resulting set of formulas is given below. \\[1ex]
Thermodynamical quantities:
\begin{flalign}
&T_* = \frac{\partial e(\rhoc, s)}{\partial s} \, , & \\
&\pi_* \equiv \frac{p(\rhoc,s)}{\rhoc} = \rhoc \frac{\partial
e(\rhoc,s)}{\partial \rhoc} \, , & \\ 
& \gammac \equiv \frac{\partial \ln p(\rhoc, s)}{\partial \ln \rhoc} \, , & \\
&\hc = e(\rhoc,s) + \pi_* \, .
\end{flalign}
Primary Poisson equations:
\begin{flalign} 
&\Delta \Uc = - 4 \pi G \rhoc \, , & \\
&\Delta \Uc_i = - 4 \pi G \rhoc w_i \, . &
\end{flalign}
$2.5$ PN quantities:\\[\bigskipamount]
$Q_{ij}^{[2]}$, see Appendix \ref{sec:appB};
\begin{flalign}
&\dot{Q}_{ij}^{[2]} = \frac{dQ_{ij}^{[2]}}{dt} \,  & \\ 
&{h_5}_{ij} = - \frac{2 G}{5} \dot{Q}_{ij}^{[2]} \, , & \\
&\Delta R = -4 \pi G \dot{Q}_{ij}^{[2]} x^i
\partial_j \rhoc \, , & \\
&U_5 = \frac{2G}{5} \left(- \dot{Q}_{ij}^{[2]} x^i \partial_j \Uc + R \right)
\, . &
\end{flalign}
1 PN quantities with radiative corrections included:
\begin{flalign}
&\alpha = 2 \Uc - \gammac \left(\frac{1}{2} w^2 + 3 \Uc\right) & \nonumber \\
 & \quad ~ \;  + \frac{1}{c^5}  \left[ 2 U_5 + 
 \gammac \left(- 3 U_5  + \wc_k \wc_l {h_5}_{kl} \right) \right] \, , & \\
&\beta = \frac{1}{2} w^2 + \hc + 3 \Uc \, , &  \\
& \delta = \frac{3}{2} w^2 + \hc + 2 \pi_* - \Uc \, , & \\ 
& \zeta_i = - \partial_j (\rhoc w_i w_j) + \rhoc \partial_i \Uc - 
\partial_i \pc \, , & \\ 
& \Delta U_2 = -4 \pi G \rhoc \delta \, , & \\
&\Delta D_i = -4 \pi G \zeta_i \, , & \\
& \Delta E_i = -4 \pi G x^i \partial_j \zeta_j \, ,  & \\
& \Delta C_i = - 4 \pi G \left(\delta^{jk} - \frac{2}{c^5} {h_5}_{jk}\right) 
x^i \partial_j (\rhoc w_k) \, , & \\
&\Apot_i = 4 \left(\delta^{ik} - \frac{2}{c^5} {h_5}_{ik}\right) \Uc_k +
 \frac{1}{2} C_i & \nonumber \\ & \qquad \; - \frac{1}{2} x^i
 \left(\delta^{jk} -  \frac{2}{c^5} {h_5}_{jk}\right) \partial_j \Uc_k \, . &
\end{flalign}
$3.5$ PN quantities:\\[\bigskipamount]
${I_2}_{ij}^{[2]}$, $Q_{ij}^{[3]}$, $Q_{ij}^{[4]}$, $Q_{ijk}^{[4]}$,
$S_{ij}^{[3]}$, see Appendix \ref{sec:appB}; \vspace*{-\smallskipamount}
\begin{flalign}
&{\dot{I}_2{}}^{[2]}_{ij} = \frac{d{I_2}^{[2]}_{ij}}{dt}\, , & \\
& \dot{Q}_{ij}^{[4]} = \frac{dQ_{ij}^{[4]}}{dt}\, , & \\
& \dot{Q}_{ijk}^{[4]} = \frac{dQ_{ijk}^{[4]}}{dt}\, , & \\
& \dot{S}_{ij}^{[4]} = \frac{dS_{ij}^{[4]}}{dt}\, , & \\
& \Delta R_2 = - 4 \pi G \bigg[
{\dot{I}_2{}}_{kl}^{[2]} x^k \partial_l \rhoc +
Q_{kl}^{[3]} x^k \left(\rhoc \partial_l \Uc +\partial_l (\rhoc
\delta)\right) 
& \nonumber \\ & \qquad \quad \; +  3 \rhoc \wc_k \wc_l Q_{kl}^{[3]}
- \frac{5}{126} \dot{Q}_{klm}^{[4]} x^k x^l \partial_m \rhoc & \nonumber \\ &
\qquad \quad \; +  Q_{kl}^{[4]} x^k 
\left(\frac{1}{2} x^l \partial_t \rhoc - 4 \rhoc \wc_l \right) & \nonumber \\ &
\qquad \quad \; + \dot{Q}_{kl}^{[4]} x^k \left(- \frac{17}{42} x^l x^m
  \partial_m \rhoc + 
\frac{11}{42} r^2 \partial_l \rhoc - \rhoc x^l \right) & \nonumber \\ & \qquad
\quad \; -
\frac{8}{9} \epsilon_{klm} \dot{S}_{mn}^{[3]} x^l x^n \partial_k \rhoc - \rhoc
R \bigg] \, , & \\
&\Delta R_i = - 4 \pi G \bigg[Q_{kl}^{[3]} x^k \partial_l 
\left(\rhoc w_i \right) - \rhoc w_k Q_{ik}^{[3]} - 
\rhoc x^k Q_{ik}^{[4]} & \nonumber \\ & \qquad \quad \; - \frac{1}{8}
\partial_t (Q_{kl}^{[3]} x^i x^k \partial_l \rhoc) \bigg] \, , & \\
&U_7 = \frac{2 G}{5} \bigg[ - {\dot{I}_2{}}_{kl}^{[2]} x^k \partial_l  \Uc +
Q_{kl}^{[3]} x^k \left(- \partial_l U_2 + 2 \Uc \partial_l \Uc \right) 
& \nonumber \\ & \qquad  \; + 
Q_{kl}^{[4]} x^k \left(- \frac{1}{2} x^l \partial_t \Uc + \Apot_l \right) 
+ \frac{5}{126} \dot{Q}_{klm}^{[4]} x^k x^l \partial_m \Uc \nonumber & \\ &
\qquad \; + \dot{Q}_{kl}^{[4]} x^k \left( \frac{17}{42} x^l x^m \partial_m \Uc
  - \frac{11}{42} r^2 \partial_l \Uc \right) \nonumber & \\ & \qquad \; -
\frac{8}{9} \epsilon_{klm} \dot{S}_{mn}^{[3]} x^l x^n \partial_k \Uc -2 \Uc R
+ R_2 \bigg] \, , & \\
& {U_{5}}_i = \frac{2 G}{5} \bigg[ - 
\frac{1}{4} Q_{kl}^{[3]} x^k \partial_l \Apot_i - 
\frac{1}{4} Q_{ik}^{[3]} \Apot_k + 
Q_{ik}^{[4]} x^k \Uc & \nonumber \\ & \qquad ~ \;  + \frac{1}{8} x^i
\partial_t R + R_i \bigg] \, , & \\ 
& {h_7}_{ij} = \frac{2 G}{5} \bigg[ - 2 {\dot{I}_2{}}_{ij}^{[2]} - 2
Q_{ij}^{[3]} \Uc 
+ \frac{5}{63} \dot{Q}_{ijk}^{[4]} x^k + \frac{2}{7} x^k x^{(i}
\dot{Q}_{j)k}^{[4]} & \nonumber \\ & \qquad \quad -
\frac{11}{42} \dot{Q}_{ij}^{[4]} r^2  - 
\frac{8}{9} \epsilon_{kl(i} \dot{S}_{j)l}^{[3]} x^k
- \frac{2}{21} \delta^{ij} x^k
x^l \dot{Q}_{kl}^{[4]} \bigg] \, .
\end{flalign}
Velocity and forces:
\begin{flalign}
&v^i = \wc_i + \frac{1}{c^2} \left(- \beta \wc_i + \Apot_i \right) -  
\frac{2}{c^5} \wc_k {h_5}_{ik} & \nonumber \\ & \qquad - \frac{1}{c^7}
\bigg(- 4 {U_5}_i + 2 \wc_k 
{h_7}_{ik} + \wc_i \left(3 U_5 - \wc_k \wc_l {h_5}_{kl} 
\right) & \nonumber \\ &  \qquad  - 2 (\beta + 2 \Uc) \wc_k {h_5}_{ik}
+ 2 \Apot_k {h_5}_{ik} \bigg) \, , & \\
&F^{\rm press}_i = - \left(1 + \frac{\alpha}{c^2}\right) 
\left(\partial_i \hc - T_* \partial_i s \right) - 
\frac{1}{c^2} \pi_* \partial_i \alpha \, , & \\
&F^{\rm grav}_i = \left(1 + \frac{\delta}{c^2} \right) \partial_i \Uc  +
\frac{1}{c^2}  
\partial_i U_2 - \frac{1}{c^2} w^j \partial_i  \Apot_j +
\frac{1}{c^5} \partial_i U_5 & \nonumber \\ & \qquad \quad \; + \frac{1}{c^7}
\bigg( \partial_i U_7 - 4 \wc_k \partial_i {U_5}_k + \wc_k \wc_l
\partial_i {h_7}_{kl} & \nonumber \\ & \qquad \quad \;  + \left(U_5 - 5 \wc_k
\wc_l {h_5}_{kl} \right) \partial_i \Uc  + \left(\delta + 2 \Uc \right)
\partial_i U_5 & \nonumber \\ & \qquad \quad \; + 2 {h_5}_{kl} \wc_k
\partial_i \Apot_l \bigg) \, . & 
\end{flalign}
Evolution system:
\begin{flalign}
&\partial_t \rhoc = - \partial_i (\rhoc v^i) \, , &\\
&\partial_t s = - v^i \partial_i s \, , &\\
&\partial_t w_i = - v^j \partial_j w_i + F^{\rm grav}_i + F^{\rm press}_i \, .
& 
\end{flalign}
Gravitational wave form $(R= \sqrt{X^i X^i}, \mathbf{N}=\mathbf{X}/R)$:
\begin{flalign}
  &h_{ij}^{\rm rad} (\mathbf{X},T) = \frac{2G}{c^4 R} P_{ijkl}(\mathbf{N})
  \left\{ I_{kl}^{(2)} + \frac{1}{3c} N_a Q_{akl}^{[3]} \right. & \nonumber \\ 
  & \left.  \qquad \qquad \quad~ + \frac{4}{3c} \epsilon_{ab(k} S^{[2]}_{l)a}
    N_b + \frac{1}{12 c^2} N_a N_b \dot{Q}_{abkl}^{[3]} \right. & \nonumber \\ 
  & \left.  \qquad \qquad \quad ~ + \frac{1}{2 c^2} \epsilon_{ab(k}
    \dot{S}^{[2]}_{l)ac} N_b N_c \right\} \, . &
\end{flalign}
\end{subequations}

\section*{Acknowledgment}

This work was supported by the EU Program ``Improving the Human Research
Potential and the Socio-Economic Knowledge Base'' (Research Training Network
Contract HPRN-CT-2000-00137). We thank M. Br\"ugmann for checks of several
formulas.

\appendix

\section{Set of formulas in the
$\mathbf{\{}\boldsymbol{\tilde{\rho}}\mathbf{,w_i,s\}}$ representation} 
\label{sec:appA}
We give here the set of equations describing the $1$ + $3.5$ PN dynamics
by means of the matter variables 
$\tilde{\rho}= \rhoc + \sigma_{(7)}$, $w_i$, and
$s$ (see section \ref{sec:reaction}). \\[1ex]
Thermodynamical quantities:
\begin{flalign}
&\tilde{T} = \frac{\partial e(\tilde{\rho}, s)}{\partial s} \, , \\
\tilde{\pi} &\equiv \frac{p(\tilde{\rho},s)}{\tilde{\rho}} = \tilde{\rho}
 \frac{\partial e(\tilde{\rho},s)}{\partial \tilde{\rho}} \, , & \\ 
&\tilde{\gamma} \equiv \frac{\partial \ln p(\tilde{\rho}, s)}{\partial \ln
\tilde{\rho}} \, , & \\
& \tilde{h} = e(\tilde{\rho},s) + \tilde{\pi} \, .
\end{flalign}
Primary Poisson equations:
\begin{flalign} 
&\Delta \tilde{U} = - 4 \pi G \tilde{\rho} \, , & \\
&\Delta \tilde{U}_i = - 4 \pi G \tilde{\rho} w_i \, . &
\end{flalign}
$2.5$ PN quantities:\\[\bigskipamount]
$Q_{ij}^{[2]}$, see Appendix \ref{sec:appB}, with the substitutions $\rhoc
\rightarrow \tilde{\rho}$, $\Uc \rightarrow \tilde{U}$, etc...;
\begin{flalign}
&\dot{Q}_{ij}^{[2]} = \frac{dQ_{ij}^{[2]}}{dt} \,  & \\
&{h_5}_{ij} = - \frac{2 G}{5} \dot{Q}_{ij}^{[2]} \, ,\\
&\Delta R = -4 \pi G \dot{Q}_{ij}^{[2]} x^i
\partial_j \tilde{\rho} \, , & \\
&U_5 = \frac{2G}{5} \left(- \dot{Q}_{ij}^{[2]} x^i \partial_j \tilde{U} + R
\right) \, . 
\end{flalign}
1 PN quantities with radiative corrections included:
\begin{flalign}
&\alpha = 2 \tilde{U} - \tilde{\gamma} \left(\frac{1}{2} w^2 + 3
 \tilde{U}\right) + \frac{2}{c^5} 
 U_5 (1 - \tilde{\gamma}) \, , & \\
&\beta = \frac{1}{2} w^2 + \tilde{h} + 3 \tilde{U} \, , & \\ 
&\delta = \frac{3}{2} w^2 + \tilde{h} + 2 \tilde{\pi} - \tilde{U} \, , & \\ 
& \zeta_i = - \partial_j (\tilde{\rho} w_i w_j) + \tilde{\rho} \partial_i
 \tilde{U} -  \partial_i \tilde{p} \, , \\
& \Delta U_2 = -4 \pi G \tilde{\rho} \delta \, , & \\
&\Delta D_i = -4 \pi G \zeta_i \, , & \\
&\Delta E_i = -4 \pi G x^i \partial_j \zeta_j \, ,  \\
& \Delta C_i = - 4 \pi G \left(\delta^{jk} - \frac{2}{c^5} {h_5}_{jk}\right) 
x^i \partial_j (\tilde{\rho} w_k) \, , & \\
&\tilde{A}_i = 4 \left(\delta^{ik} - \frac{2}{c^5} {h_5}_{ik}\right)
 \tilde{U}_k + 
 \frac{1}{2} C_i & \nonumber \\ & \qquad \; - \frac{1}{2} x^i
 \left(\delta^{jk} -  \frac{2}{c^5} {h_5}_{jk}\right) \partial_j \tilde{U}_k
 \, . & 
\end{flalign}
$3.5$ PN quantities:\\[\bigskipamount]
${I_2}_{ij}^{[2]}$, $Q_{ij}^{[3]}$, $Q_{ij}^{[4]}$, $Q_{ijk}^{[4]}$,
$S_{ij}^{[3]}$, see Appendix \ref{sec:appB}, with the substitutions $\rhoc
\rightarrow \tilde{\rho}$, $\Uc \rightarrow \tilde{U}$, etc...;
\vspace*{-\smallskipamount} 
\begin{flalign}
&{\dot{I}_2{}}^{[2]}_{ij} = \frac{d{I_2}^{[2]}_{ij}}{dt}\, & \\
&\dot{Q}_{ij}^{[4]} = \frac{dQ_{ij}^{[4]}}{dt}\, , & \\
&\dot{Q}_{ijk}^{[4]} = \frac{dQ_{ijk}^{[4]}}{dt}\, , & \\
&\dot{S}_{ij}^{[4]} = \frac{dS_{ij}^{[4]}}{dt}\, , & \\
& \theta = \frac{2G}{5} \left[Q_{ij}^{[3]} (x^i \partial_j \tilde{U} -
w_i w_j)- R \right] \, , & \\
&\Delta R_2 = - 4 \pi G \bigg[
{\dot{I}_2{}}_{kl}^{[2]} x^k \partial_l \tilde{\rho} +
Q_{kl}^{[3]} x^k \partial_l (\tilde{\rho} \delta) & \nonumber \\ & \qquad \quad
\; +  4 \tilde{\rho} \wc_k \wc_l Q_{kl}^{[3]}  -
\frac{5}{126} \dot{Q}_{klm}^{[4]} x^k x^l \partial_m \tilde{\rho} & \nonumber
\\ & \qquad \quad \; + 
Q_{kl}^{[4]} x^k \left(\frac{1}{2} x^l \partial_t \tilde{\rho} - 4
  \tilde{\rho} \wc_l \right) \nonumber \\ 
& \qquad \quad \; + \dot{Q}_{kl}^{[4]} x^k \left(- \frac{17}{42} x^l x^m
  \partial_m 
\tilde{\rho} + \frac{11}{42} r^2 \partial_l \tilde{\rho} - 
\tilde{\rho} x^l \right)& \nonumber \\ & \qquad  \quad \;  -
\frac{8}{9} \epsilon_{klm} \dot{S}_{mn}^{[3]} x^l x^n \partial_k \tilde{\rho}
\bigg] \, , & \\ 
&\Delta R_i = - 4 \pi G \bigg[Q_{kl}^{[3]} x^k \partial_l 
\left(\tilde{\rho} w_i \right) - \tilde{\rho} w_k Q_{ik}^{[3]} - 
\tilde{\rho} x^k Q_{ik}^{[4]} & \nonumber \\ & \qquad \quad \; - \frac{1}{8}
\partial_t (Q_{kl}^{[3]} x^i x^k 
\partial_l \tilde{\rho}) \bigg] \, , & \\
&U_7 = \frac{2 G}{5} \bigg[ - {\dot{I}_2{}}_{kl}^{[2]} x^k \partial_l
	\tilde{U} + 
Q_{kl}^{[3]} x^k \left(- \partial_l U_2 + 2 \tilde{U} \partial_l \tilde{U}
\right)  & \nonumber \\ & \qquad \; + 
Q_{kl}^{[4]} x^k \left(- \frac{1}{2} x^l \partial_t \tilde{U} + 
\tilde{A}_l \right) \nonumber & \\ & \qquad \; 
+ \frac{5}{126} \dot{Q}_{klm}^{[4]} x^k x^l \partial_m \tilde{U} \nonumber &
\\ & \qquad \; + 
\dot{Q}_{kl}^{[4]} x^k \left( \frac{17}{42} x^l x^m \partial_m \tilde{U} -
\frac{11}{42} r^2 \partial_l \tilde{U}  \right) & \nonumber \\ & \qquad \; -
\frac{8}{9} \epsilon_{klm} \dot{S}_{mn}^{[3]} x^l x^n \partial_k \tilde{U} -2
\tilde{U} R + R_2 \bigg] \, , & \\
& {U_{5}}_i = \frac{2 G}{5} \bigg[ - 
\frac{1}{4} Q_{kl}^{[3]} x^k \partial_l \tilde{A}_i - 
\frac{1}{4} Q_{ik}^{[3]} \tilde{A}_k + 
Q_{ik}^{[4]} x^k \tilde{U} \nonumber & \\ & \qquad \quad + \frac{1}{8} x^i
\partial_t R + R_i \bigg] \, , & \\ 
& {h_7}_{ij} = \frac{2 G}{5} \bigg[ - 2 {\dot{I}_2{}}_{ij}^{[2]} - 2
Q_{ij}^{[3]} \tilde{U} + \frac{5}{63} \dot{Q}_{ijk}^{[4]} x^k + \frac{2}{7}
x^k x^{(i} \dot{Q}_{j)k}^{[4]} & \nonumber \\ & \qquad \quad -
\frac{11}{42} \dot{Q}_{ij}^{[4]} r^2  - 
\frac{8}{9} \epsilon_{kl(i} \dot{S}_{j)l}^{[3]} x^k
- \frac{2}{21} \delta^{ij} x^k
x^l \dot{Q}_{kl}^{[4]} \bigg] \, .
\end{flalign}
Baryonic density, velocity and forces:
\begin{flalign}
&\rhoc = \tilde{\rho} \left(1 - \frac{\theta}{c^7} \right) \, , & \\
&v^i = \wc_i + \frac{1}{c^2} \left(- \beta \wc_i + \tilde{A}_i \right) -  
\frac{2}{c^5} \wc_k {h_5}_{ik} & \nonumber \\ & \qquad - \frac{1}{c^7} \bigg(-
4 {U_5}_i + 2 \wc_k 
{h_7}_{ik} + \wc_i \left(3 U_5 - \wc_k \wc_l {h_5}_{kl}
\right) & \nonumber \\ & \qquad - 2 (\beta + 2 \tilde{U}) \wc_k {h_5}_{ik} + 2
\tilde{A}_k {h_5}_{ik} \bigg) \, , & \\
&F^{\rm press}_i = - \left[1 + \frac{\alpha}{c^2} + \frac{\theta}{c^7} \right] 
\left(\partial_i \tilde{h} - \tilde{T} \partial_i s \right) - 
\frac{1}{c^2} \tilde{\pi} \partial_i \alpha \, , & \\
&F^{\rm grav}_i = \left(1 + \frac{\delta}{c^2} \right) \partial_i \tilde{U}  +
\frac{1}{c^2}  
\partial_i U_2 - \frac{1}{c^2} w^j \partial_i  \tilde{A}_j +
\frac{1}{c^5} \partial_i U_5 & \nonumber \\ & \qquad \quad \;  + \frac{1}{c^7}
\bigg( \partial_i U_7 - 4 \wc_k \partial_i {U_5}_k + \wc_k \wc_l
\partial_i {h_7}_{kl} & \nonumber \\ & \qquad \quad \; + \left(U_5 - 5 \wc_k
  \wc_l {h_5}_{kl} \right) \partial_i \tilde{U} + \left(\delta + 2
  \tilde{U} \right) \partial_i U_5  & \nonumber \\ & \qquad \quad \; + 2
{h_5}_{kl} \wc_k \partial_i \tilde{A}_l \bigg) \, . & 
\end{flalign}
Evolution system:
\begin{flalign}
& \partial_t \rhoc = - \partial_i (\rhoc v^i) \, , & \\
& \partial_t s = - v^i \partial_i s \, , & \\
& \partial_t w_i = - v^j \partial_j w_i + F^{\rm grav}_i + F^{\rm press}_i \, .
&
\end{flalign}

\begin{widetext}
\section{Time derivatives of relevant multipole moments} \label{sec:appB}

This Appendix presents the full explicit expressions for the time derivatives
of the multipole moments used in the present formalism.
\begin{subequations} 
\begin{align} 
I^{(1)}_{ij} =& Q_{ij}^{[1]} + \frac{1}{c^2} {I_2}^{[1]}_{ij} +
\cO\left(\frac{1}{c^4} \right) \\  
=&  \textrm{STF} \integ \bigg \{2 \rhoc x^i \wc_j  +
\frac{1}{c^2} \bigg[ \rhoc  
\big(2 \Apot_i x^j + \frac{11}{7} r^2 \wc_j \partial_i \Uc 
+ \frac{10}{21} x^p \wc_i \wc_j \wc_p \big) 
+ \frac{22}{21} \pc r^2 \partial_i \wc_j \nonumber \\ & \quad
+ \rhoc x^i \big[\frac{11}{21} r^2 \wc_p \partial_{jp} \Uc 
- \frac{11}{21} r^2 \partial_{jp} \Uic_p -8 \Uc \wc_j -\frac{4}{21} \wc_j
\wc^2 \big] \nonumber \\
& \quad + x^i x^p \big[\rhoc(- \frac{46}{21} \wc_j \partial_p \Uc 
+ \frac{10}{21}
\wc_p \partial_j \Uc \big) -\frac{4}{7} \pc \big(\partial_p \wc_j 
+ \partial_j \wc_p \big) \big] + x^i x^j \big[\rhoc \big(\partial_p \Uic_p
-\frac{3}{7} \wc_p \partial_p\Uc 
\big) + \frac{8}{21} \pc \partial_p \wc_p \big] \nonumber \\ 
& \quad +\frac{17}{21} \rhoc  x^i x^j x^p \big[\partial_{pq} \Uic_q 
- \wc_q \partial_{pq} \Uc \big] \bigg] \bigg\} \,\dx+ \cO\left(\frac{1}{c^4}
\right)\, , \\ 
I_{ij}^{(2)} &= Q_{ij}^{[2]} + \frac{1}{c^2} {I_2}_{ij}^{[2]} +
\cO\left(\frac{1}{c^4} \right) = \textrm{STF}
\integ \bigg\{ 2 \rhoc x^i \partial_j \Uc + 2 
\rhoc \wc_i \wc_j \nonumber \\ & + \frac{1}{c^2} \bigg[\rhoc \big( 8 D_i
x^j + E_i x^j + \frac{11}{7} r^2 \partial_i \Uc \partial_j \Uc + 4 \Apot_i
\wc_j - \frac{44}{21} r^2 \wc_j \partial_{ip} \Uc_p - 14
\Uc \wc_i \wc_j \nonumber \\ & \qquad + \frac{44}{21} r^2 \wc_j
\wc_p\partial_{ip} \Uc - \frac{5}{7} 
\wc_i \wc_j \wc^2 \big) + \pc \big(- \frac{22}{21} r^2
\partial_i \wc_p \partial_p \wc_j + \frac{22}{21} (1-\gammac) r^2
\partial_i \wc_j \partial_p \wc_p + \frac{22}{7} r^2 \partial_{ij} \Uc 
\big) \nonumber \\ 
& \qquad - 2 \hc \rhoc \wc_i \wc_j + 2 \pc \wc_i \wc_j + \frac{22}{21} r^2
\partial_i \hc \partial_j \pc - \frac{22}{21} r^2 T_* \partial_i s \partial_j
\pc \nonumber \\ & \qquad + x^p
\big[\rhoc \big(- \frac{12}{7} \wc_i \wc_j \partial_p \Uc + \frac{32}{7}
\wc_j \wc_p \partial_i \Uc \big) + 
\pc \big( \frac{8}{21} \wc_i \partial_p \wc_j  
+ \frac{8}{21} \wc_j \partial_i \wc_p +
\frac{64}{21} \wc_p \partial_i \wc_j \big) \big]
\nonumber \\ & \qquad +
x^i \big[ \rhoc \big(2 \partial_j U_2  -
10 \Uc \partial_j \Uc + 10 \wc_j \partial_p \Uc_p  -
2 \wc_p \partial_j \Apot_p + 2 \wc_p \partial_p \Apot_j + \frac{23}{7}
\wc^2 \partial_j \Uc  -
\frac{80}{7} \wc_j \wc_p \partial_p \Uc \big) \nonumber \\ & \qquad \quad \;
\, + \pc \big( - 2 \partial_j \Uc + 
\frac{8}{21} \wc_j \partial_p \wc_p  -
\frac{20}{21} \wc_p \partial_j \wc_p -
\frac{20}{21} \wc_p \partial_p \wc_j \big) + 
2 \hc \rhoc \partial_j \Uc \big] \nonumber \\ & 
\qquad + x^i x^p \big[\rhoc \big(- \frac{12}{7} \partial_j \Uc \partial_p \Uc
+  \frac{80}{21} \wc_j \partial_{pq} \Uc_q - 
\frac{32}{21} \wc_p \partial_{jq} \Uc_q - 
\frac{80}{21}  \wc_j \wc_q \partial_{pq} \Uc + 
\frac{32}{21}  \wc_p \wc_q \partial_{jq} \Uc \big) 
\nonumber \\ & \qquad \qquad \; + \pc \big(
\frac{4}{7}  \partial_p \wc_q \partial_q \wc_j +  
\frac{4}{7}  \partial_j \wc_q \partial_q \wc_p - 
\frac{4}{7} (1-\gammac) \partial_p \wc_j \partial_q \wc_q - 
\frac{4}{7} (1-\gammac) \partial_j \wc_p \partial_q \wc_q - 
\frac{24}{7}  \partial_{jp} \Uc \big) \nonumber \\ & \qquad \qquad \quad -
\frac{8}{7} \partial_j \pc \partial_p \hc +
\frac{8}{7} T_* \partial_j \pc \partial_p s \big] \nonumber \\ &
\qquad + x^i x^j \big[ \rhoc \big(-\frac{3}{7} \partial_p \Uc \partial_p \Uc +
\frac{47}{21}  \wc_q \partial_{pq} \Uc_p - 
\frac{26}{21}  \wc_p \wc_q \partial_{pq} \Uc \big) \nonumber \\ & \qquad \qquad
\; + \pc \big(
\frac{8}{21} (1-\gammac) (\partial_p \wc_p)^2 - 
\frac{8}{21} \partial_q \wc_p \partial_p \wc_q  - 
\frac{6}{7} (-4 \pi G \rhoc)  + \frac{8}{21} \partial_p \hc \partial_p \pc - 
\frac{8}{21} T_* \partial_p s \partial_p \pc \big] \nonumber \\ & 
\qquad + r^2 x^i \big[ \rhoc \big(
\frac{11}{21} \partial_p \Uc \partial_{jp} \Uc - 
\frac{11}{21} \partial_{jp} D_p  - 
\frac{22}{21} \wc_q \partial_{jpq} \Uc_p + 
\frac{11}{21} \wc_p \wc_q \partial_{jpq} \Uc \big) +
\frac{11}{21} \pc \partial_j (- 4 \pi G \rhoc) \big] \nonumber \\ & \qquad + 
x^i x^j x^p \big[\rhoc \big(-
\frac{17}{21} \partial_q \Uc \partial_{pq} \Uc + 
\frac{17}{21}  \partial_{pq} D_q + 
\frac{34}{21}  \wc_r \partial_{pqr} \Uc_q \nonumber \\ & \qquad \qquad  - 
\frac{17}{21}  \wc_q \wc_r \partial_{pqr} \Uc \big) - 
\frac{17}{21}  \pc \partial_{p} (- 4 \pi G \rhoc)\big]\bigg] 
\bigg\} \,\dx + \cO\left(\frac{1}{c^4} \right)\, , \\
Q^{[3]}_{ij} &= \textrm{STF} \integ \Big\{4 \pc \partial_i \wc_j + 6 \rhoc
\wc_j \partial_i \Uc  
+ 2 \rhoc x^i \big[\wc_p \partial_{jp} \Uc - \partial_{jp} \Uic_p \big] \Big
\} \,\dx \, , \\
Q^{[4]}_{ij} &= \textrm{STF} \integ \Big\{ 4 \rhoc (\partial_i \hc - T_*
\partial_i s) (\partial_j \hc - T_* \partial_j s) + 2 \rhoc \big(3 \partial_i
\Uc \partial_j \Uc  
- 4 \wc_j \partial_{ip} \Uic_p + 4 \wc_j \wc_p \partial_{ip} \Uc \big)
\nonumber \\ 
& \quad + 4 \pc \big(- \partial_i \wc_p \partial_p \wc_j 
+ \partial_i \wc_j \partial_p \wc_p  
-  \gammac \partial_i \wc_j \partial_p \wc_p + 3\partial_{ij} \Uc  \nonumber
\\  & \quad +  x^i \big[2 \rhoc \big( \partial_{jp} \Uc\partial_p \Uc 
- \partial_{jp} \Di_p - 2 \wc_q \partial_{jpq} \Uic_p 
+ \wc_p \wc_q \partial_{jpq} \Uc \big)
+2 \pc \partial_{j} (-4\pi G \rhoc) \big] \Big\} \,\dx \, , \\
Q^{[1]}_{ijk} &= \textrm{STF} \integ \Big\{ 3 \rhoc x^i x^j \wc_k \Big\}
\,\dx \, , \\ 
Q^{[2]}_{ijk} & = \textrm{STF} \integ \Big\{ 
3 \rhoc  x^i x^j \partial_k \Uc +  
6 \rhoc x^i \wc_j \wc_k \Big\} \,\dx \, ,\\
Q^{[3]}_{ijk} & = \textrm{STF} \integ \Big\{ 6 \rhoc \wc_i \wc_j \wc_k + 
x^i \big[18 \rhoc \wc_k \partial_j \Uc +12 \pc \partial_j \wc_k \big]
+ 3 \rhoc  x^i x^j \big[-\partial_{kp} \Uic_p 
+ \wc_p \partial_{kp} \Uc \big] \Big\} \,\dx \, ,\\
Q^{[4]}_{ijk} & = \textrm{STF} \integ \Big\{36 \rhoc \wc_j \wc_k \partial_i
\Uc + 48 \pc \wc_j \partial_i \wc_k +  x^i \big[12 \rhoc (\partial_j \hc - T_*
\partial_j s) (\partial_k \hc - T_* \partial_k s)
 \nonumber \\ & \qquad + \rhoc \big(18 \partial_j \Uc \partial_k \Uc 
 - 24\wc_k \partial_{jp} \Uic_p + 24 \wc_k \wc_p \partial_{jp} \Uc
\big)\nonumber \\ 
& \qquad +\pc \big(-12 \partial_j \wc_p \partial_p \wc_k 
+12\partial_j \wc_k \partial_p \wc_p
- 12 \gammac \partial_j \wc_k \partial_p \wc_p 
+ 36 \partial_{jk} \Uc \big) \big]\nonumber \\ 
& + x^i x^j \big[\rhoc \big(3 \partial_p \Uc \partial_{kp} \Uc -3
\partial_{kp} \Di_p - 6 \wc_q \partial_{kpq}
\Uic_p  
+ 3 \wc_p \wc_q \partial _{kpq} \Uc \big) \big]+ 3 \pc \partial_{k} (-4\pi G
\rhoc) \Big\} \,\dx \, ,\\
Q^{[1]}_{ijkl} & = \textrm{STF} \integ \Big\{ 4 \rhoc x^i x^j x^k \wc_l
\Big\} \dx \, , \\
Q^{[2]}_{ijkl} & = \textrm{STF} \integ \Big\{ 4 \rhoc x^i x^j x^k \partial_l
\Uc + 12 \rhoc x^i x^j \wc_k \wc_l \Big\} \,\dx \, , \\
Q^{[3]}_{ijkl} &= \textrm{STF} \integ \Big\{ 24 \rhoc x^i \wc_j \wc_k \wc_{l} 
+ x^i x^j \big(36 \rhoc \wc_l \partial_k \Uc + 24 \pc \partial_k \wc_l \big)
+ 4 \rhoc x^i x^j x^k \big(- \partial_{lp} \Uc_p+ 
 \wc_p \partial_{lp} \Uc \big) \Big\} \,\dx \, ,\\
S^{[1]}_{ij} & = \textrm{STF}\integ \Epsilon_{pqi}  \Big\{ \rhoc  x^j x^p
\partial_q \Uc + \rhoc x^p \wc_j \wc_q \bigg \} \,\dx \, , \\
S^{[2]}_{ij} & =\textrm{STF} \integ \Epsilon_{pqi} \Big\{ x^p
\big[\rhoc 
\wc_q \partial_j \Uc + \pc \partial_j \wc_q \big] \nonumber \\
& \quad + \Big[ x^p (2 \rhoc \wc_j \partial_q \Uc +\pc
\partial_q \wc_j) + x^j \rhoc \wc_p \partial_q \Uc 
+ \rhoc x^j x^p (\wc_r \partial_{qr} \Uc - \partial_{qr} \Uic_r 
) \Big] \Big\} \,\dx \, , \\
S^{[3]}_{ij} &= \textrm{STF} \integ \Epsilon_{pqi}\Big\{ \big[-\pc \wc_q 
\partial_j \wc_p 
+ \rhoc x^p \big[- \wc_q \partial_{jr} \Uic_r 
+ \wc_q \wc_r \partial_{jr} \Uc \big]\nonumber \\
& \qquad +\pc x^p \big[- \partial_j \wc_r \partial_r \wc_q
+\partial_j \wc_q \partial_r \wc_r 
- \gammac \partial_j \wc_q \partial_r \wc_r \big] \big]\nonumber \\ 
& \qquad + \big[ 2 \rhoc x^p (\partial_j \hc - 
T_* \partial_j s) (\partial_q \hc - T_* \partial_q s) + \pc \wc_p \partial_q
\wc_j + 3 \rhoc \wc_j \wc_p \partial_q \Uc \nonumber \\ & \qquad   
+ 3 \rhoc x^p \big[\partial_j \Uc \partial_q \Uc
-\wc_j \partial_{qr} \Uic_r + \wc_j \wc_r \partial_{qr} \Uc \big]
+ \pc x^p \big[\partial_q \wc_j \partial_r \wc_r
-\partial_r \wc_j \partial_q \wc_r
- \gammac\partial_q \wc_j \partial_r \wc_r + 6 \partial_{jq} \Uc  \big]
\nonumber \\ 
& \qquad + 2 \rhoc x^j \big[\wc_p \wc_r
\partial_{qr} \Uc - \wc_p \partial_{qr} \Uic_r \big] \nonumber \\
& \qquad + x^j x^p \big[\rhoc \big(\partial_{qr} \Uc \partial_r \Uc
- \partial_{qr} \Di_r 
- 2\wc_s \partial_{qrs} \Uic_r 
+\wc_r \wc_s \partial_{qrs} \Uc \big)+\pc \partial_{q} (-4\pi G \rhoc)
\big] \big] \Big\} \,\dx \, , \\
S^{[1]}_{ijk} & = \textrm{STF} \integ \Epsilon_{pqi}\Big\{ \rhoc  x^j x^k x^p
\partial_q \Uc
+ 2 \rhoc  x^j x^p \wc_k \wc_q \Big\} \,\dx \, , \\
S^{[2]}_{ijk} &= \textrm{STF}  \integ \Epsilon_{pqi} \Big\{ 2 \rhoc x^p \wc_j 
\wc_k \wc_q
+ \rhoc x^j x^k \wc_p \partial_q\Uc 
+  x^j x^p \big[2 \rhoc  \big(2 \wc_k \partial_q \Uc 
+ \wc_q \partial_k \Uc \big) + 2 \pc \big(
\partial_q \wc_k +\partial_k \wc_q \big)  \big]\nonumber \\[-0.1cm]
& \quad - \rhoc x^j x^k x^p \big[\partial_{qr}
\Uic_r - \wc_r \partial_{qr} \Uc \big] \Big\} \,\dx \, . \\
\end{align}
\end{subequations}
\end{widetext}


\begin{thebibliography}{16}
\expandafter\ifx\csname natexlab\endcsname\relax\def\natexlab#1{#1}\fi
\expandafter\ifx\csname bibnamefont\endcsname\relax
  \def\bibnamefont#1{#1}\fi
\expandafter\ifx\csname bibfnamefont\endcsname\relax
  \def\bibfnamefont#1{#1}\fi
\expandafter\ifx\csname citenamefont\endcsname\relax
  \def\citenamefont#1{#1}\fi
\expandafter\ifx\csname url\endcsname\relax
  \def\url#1{\texttt{#1}}\fi
\expandafter\ifx\csname urlprefix\endcsname\relax\def\urlprefix{URL }\fi
\providecommand{\bibinfo}[2]{#2}
\providecommand{\eprint}[2][]{\url{#2}}

\bibitem[{\citenamefont{Blanchet}(1997{\natexlab{a}})}]{B97}
\bibinfo{author}{\bibfnamefont{L.}~\bibnamefont{Blanchet}},
  \bibinfo{journal}{Phys. Rev. D} \textbf{\bibinfo{volume}{55}},
  \bibinfo{pages}{714} (\bibinfo{year}{1997}{\natexlab{a}}).

\bibitem[{\citenamefont{Sch\"afer}(1983)}]{S83}
\bibinfo{author}{\bibfnamefont{G.}~\bibnamefont{Sch\"afer}},
  \bibinfo{journal}{Lett. Nuovo Cim.} \textbf{\bibinfo{volume}{36}},
  \bibinfo{pages}{105} (\bibinfo{year}{1983}).

\bibitem[{\citenamefont{Damour and Sch\"afer}(1985)}]{DS85}
\bibinfo{author}{\bibfnamefont{T.}~\bibnamefont{Damour}} \bibnamefont{and}
  \bibinfo{author}{\bibfnamefont{G.}~\bibnamefont{Sch\"afer}},
  \bibinfo{journal}{Gen. Rel. Grav.} \textbf{\bibinfo{volume}{17}},
  \bibinfo{pages}{879} (\bibinfo{year}{1985}).

\bibitem[{\citenamefont{Rezzolla et~al.}(1999)\citenamefont{Rezzolla, Shibata,
  Asada, Baumgarte, and Shapiro}}]{RSABS99}
\bibinfo{author}{\bibfnamefont{L.}~\bibnamefont{Rezzolla}},
  \bibinfo{author}{\bibfnamefont{M.}~\bibnamefont{Shibata}},
  \bibinfo{author}{\bibfnamefont{H.}~\bibnamefont{Asada}},
  \bibinfo{author}{\bibfnamefont{T.~W.} \bibnamefont{Baumgarte}},
  \bibnamefont{and} \bibinfo{author}{\bibfnamefont{S.~L.}
  \bibnamefont{Shapiro}}, \bibinfo{journal}{Astrophys. J.}
  \textbf{\bibinfo{volume}{525}}, \bibinfo{pages}{935} (\bibinfo{year}{1999}).

\bibitem[{\citenamefont{Blanchet et~al.}(1990)\citenamefont{Blanchet, Damour,
  and Sch{\"a}fer}}]{BDS90}
\bibinfo{author}{\bibfnamefont{L.}~\bibnamefont{Blanchet}},
  \bibinfo{author}{\bibfnamefont{T.}~\bibnamefont{Damour}}, \bibnamefont{and}
  \bibinfo{author}{\bibfnamefont{G.}~\bibnamefont{Sch{\"a}fer}},
  \bibinfo{journal}{Mon. Not. R. astr. Soc.} \textbf{\bibinfo{volume}{242}},
  \bibinfo{pages}{289} (\bibinfo{year}{1990}).

\bibitem[{\citenamefont{Oohara and Nakamura}(1997)}]{ON97}
\bibinfo{author}{\bibfnamefont{K.}~\bibnamefont{Oohara}} \bibnamefont{and}
  \bibinfo{author}{\bibfnamefont{T.}~\bibnamefont{Nakamura}}, in
  \emph{\bibinfo{booktitle}{Relativistic Gravitation and Gravitational
  Radiation}}, edited by \bibinfo{editor}{\bibfnamefont{J.-A.}
  \bibnamefont{Marck}} \bibnamefont{and} \bibinfo{editor}{\bibfnamefont{J.-P.}
  \bibnamefont{Lasota}}, \bibinfo{organization}{Les Houches}
  (\bibinfo{publisher}{Cambridge University Press}, \bibinfo{year}{1997}).

\bibitem[{\citenamefont{Jones et~al.}()\citenamefont{Jones, Andersson, and
  Gourgoulhon}}]{JAG03}
\bibinfo{author}{\bibfnamefont{D.~I.} \bibnamefont{Jones}},
  \bibinfo{author}{\bibfnamefont{N.}~\bibnamefont{Andersson}},
  \bibnamefont{and}
  \bibinfo{author}{\bibfnamefont{E.}~\bibnamefont{Gourgoulhon}},
  \bibinfo{note}{in preparation}.

\bibitem[{\citenamefont{Stergioulas}(1998)}]{S98}
\bibinfo{author}{\bibfnamefont{N.}~\bibnamefont{Stergioulas}},
  \bibinfo{journal}{Living Reviews} \textbf{\bibinfo{volume}{1}},
  \bibinfo{pages}{1998-8} (\bibinfo{year}{1998}).

\bibitem[{\citenamefont{Chandrasekhar}(1970)}]{C70}
\bibinfo{author}{\bibfnamefont{S.}~\bibnamefont{Chandrasekhar}},
  \bibinfo{journal}{Phys. Rev. Lett.} \textbf{\bibinfo{volume}{24}},
  \bibinfo{pages}{611} (\bibinfo{year}{1970}).

\bibitem[{\citenamefont{Friedman and Schutz}(1978)}]{FS78}
\bibinfo{author}{\bibfnamefont{J.}~\bibnamefont{Friedman}} \bibnamefont{and}
  \bibinfo{author}{\bibfnamefont{B.}~\bibnamefont{Schutz}},
  \bibinfo{journal}{Astrophys. J.} \textbf{\bibinfo{volume}{222}},
  \bibinfo{pages}{281} (\bibinfo{year}{1978}).

\bibitem[{\citenamefont{Dimmelmeier et~al.}(2002)\citenamefont{Dimmelmeier,
  Font, and M\"uller}}]{DFM02a}
\bibinfo{author}{\bibfnamefont{H.}~\bibnamefont{Dimmelmeier}},
  \bibinfo{author}{\bibfnamefont{J.~A.} \bibnamefont{Font}}, \bibnamefont{and}
  \bibinfo{author}{\bibfnamefont{E.}~\bibnamefont{M\"uller}},
  \bibinfo{journal}{Astron. Astrophys.} \textbf{\bibinfo{volume}{388}},
  \bibinfo{pages}{917} (\bibinfo{year}{2002}).

\bibitem[{\citenamefont{Shibata}(2003)}]{S02}
\bibinfo{author}{\bibfnamefont{M.}~\bibnamefont{Shibata}},
  \bibinfo{journal}{Phys. Rev. D} \textbf{\bibinfo{volume}{67}},
  \bibinfo{pages}{024033} (\bibinfo{year}{2003}).

\bibitem[{\citenamefont{Schwartz}(1992)}]{inSchwartz92}
\bibinfo{author}{\bibfnamefont{L.}~\bibnamefont{Schwartz}},
  \emph{\bibinfo{title}{Analyse}} (\bibinfo{publisher}{Hermann},
  \bibinfo{address}{Paris}, \bibinfo{year}{1992}), vol.~\bibinfo{volume}{2}, p.
  \bibinfo{pages}{155}.

\bibitem[{\citenamefont{Blanchet}(1997{\natexlab{b}})}]{BH97}
\bibinfo{author}{\bibfnamefont{L.}~\bibnamefont{Blanchet}}, in
  \emph{\bibinfo{booktitle}{Relativistic Gravitation and Gravitational
  Radiation}}, edited by \bibinfo{editor}{\bibfnamefont{J.-A.}
  \bibnamefont{Marck}} \bibnamefont{and} \bibinfo{editor}{\bibfnamefont{J.-P.}
  \bibnamefont{Lasota}}, \bibinfo{organization}{Les Houches School of Physics}
  (\bibinfo{publisher}{Cambridge University Press},
  \bibinfo{address}{Cambridge}, \bibinfo{year}{1997}{\natexlab{b}}),
  p.~\bibinfo{pages}{33}.

\bibitem[{\citenamefont{Arnowitt et~al.}(1962)\citenamefont{Arnowitt, Deser,
  and Misner}}]{ADM62}
\bibinfo{author}{\bibfnamefont{R.}~\bibnamefont{Arnowitt}},
  \bibinfo{author}{\bibfnamefont{S.}~\bibnamefont{Deser}}, \bibnamefont{and}
  \bibinfo{author}{\bibfnamefont{C.~W.} \bibnamefont{Misner}}, in
  \emph{\bibinfo{booktitle}{Gravitation: An Introduction to Current Research}},
  edited by \bibinfo{editor}{\bibfnamefont{L.}~\bibnamefont{Witten}}
  (\bibinfo{publisher}{Wiley}, \bibinfo{address}{New York},
  \bibinfo{year}{1962}), chap.~\bibinfo{chapter}{7}.

\bibitem[{\citenamefont{Holm}(1985)}]{H85}
\bibinfo{author}{\bibfnamefont{D.~D.} \bibnamefont{Holm}},
  \bibinfo{journal}{Physica} \textbf{\bibinfo{volume}{17D}}, \bibinfo{pages}{1}
  (\bibinfo{year}{1985}).

\end{thebibliography}

\end{document}